\documentclass[preprint,aps,nofootinbib]{revtex4}
\usepackage{amsmath,amssymb}
\usepackage{graphicx}
\usepackage{subfigure}
\usepackage{pdflscape}
\usepackage{float}
\usepackage{color}
\usepackage{hyperref} 
\usepackage{slashed}
\usepackage{footnote}
\usepackage{multirow}
\usepackage{wrapfig}
\usepackage{textcomp}
\usepackage{amsfonts}
\usepackage{epstopdf}
\usepackage{epsfig}

\newcommand{\beq}{\begin{equation}}
\newcommand{\eeq}{\end{equation}}
\newcommand{\Lag}{\mathcal{L}}
\newcommand{\Tr}{\text{Tr}}
\newcommand{\SM}{\text{SM}}
\newcommand{\GeV}{\, \text{GeV}}
\newcommand{\TeV}{\, \text{TeV}}
\newcommand{\Det}{\, \text{Det}}
\newcommand{\be}{\begin{equation}}
\newcommand{\ee}{\end{equation}}
\newcommand{\bea}{\begin{eqnarray}}
\newcommand{\eea}{\end{eqnarray}}

\def\mF{\mathcal{F}}
\def\mG{\mathcal{G}}

\def\mL{\mathcal{L}}

\def\mO{\mathcal{O}}

\def\mT{\mathcal{T}}

\def\mX{\mathcal{X}}

\newcommand{\Eq}[1]{Eq.~(\ref{#1})}


\begin{document}

\title{Modification of Higgs Couplings in Minimal Composite Models }
\vspace*{1cm}

\author{\vspace{1cm}Da Liu$^{a}$, Ian Low$^{a,b}$ and Carlos E. M. Wagner$^{a,c,d}$}

\affiliation{
\vspace*{.5cm}
\mbox{$^a$\,High Energy Physics Division, Argonne National Laboratory, Argonne, IL 60439}\\
\mbox{$^b$\,Department of Physics and Astronomy, Northwestern University, Evanston, IL 60208}\\
\mbox{$^c$\,Enrico Fermi Institute, University of Chicago, Chicago, IL 60637}\\
\mbox{$^d$\,Kavli Institute for Cosmological Physics, University of Chicago, Chicago, IL 60637}\\
\vspace*{1cm}
}

\begin{abstract}
\vspace*{.5cm}
We present a comprehensive study of the modifications of Higgs couplings in the $SO(5)/SO(4)$ minimal composite model. We  focus on three couplings of central importance to Higgs phenomenology at the LHC: the couplings to top and bottom quarks and the coupling to two gluons. 
We consider three possible embeddings of the fermionic partners in $\mathbf{5}$, $\mathbf{10}$ and $\mathbf{14}$ of $SO(5)$ and find
 $t\bar{t}h$ and $b\bar{b}h$ couplings to be always suppressed in $\mathbf{5}$ and $\mathbf{10}$, while in $\bold{14}$ they can be either enhanced or suppressed. Assuming partial compositeness, we analyze the interplay between the $t\bar{t}h$ coupling and the top sector contribution to the Coleman-Weinberg potential for the Higgs boson,  and  the correlation between $t\bar{t}h$ and $ggh$ couplings. In particular, if the electroweak symmetry breaking is triggered radiatively by the top sector, we demonstrate that the ratio of the $t\bar{t}h$ coupling in composite Higgs models over the Standard Model expectation is preferred to be less than the corresponding ratio of the $ggh$ coupling. 
\end{abstract}

\maketitle

\section{Introduction}

The discovery of the Higgs boson at the LHC~\cite{Aad:2012tfa,Chatrchyan:2012xdj} has led to a new era in particle physics. 
The Standard Model (SM) has
been validated as the proper low energy effective theory description of the interactions between the known fundamental
particles. The Higgs boson production and decay rates seem to be in good agreement with those predicted in the SM~\cite{Khachatryan:2016vau},
suggesting that the mass generation proceeds from the Higgs mechanism, with the recently discovered Higgs being
its observable consequence.  The current precision of the Higgs rate measurements, however, leaves some room for
departures from the simple SM picture. In particular, data collected at the LHC have only started to probe Higgs couplings with the third generation quarks, which will play a central role in future Higgs measurements, while couplings to fermions in the first two generations remain a challenge. Therefore,
it is interesting to study what would be the possible consequences of the deviations
of the third generation quark couplings to the Higgs boson  and, in particular, what kind of high energy models can
accommodate such deviations in a natural way. 

A departure from the SM description is to be expected in any model that leads to the breakdown of the electroweak
symmetry in a natural way.  This could be achieved in models in which the Higgs boson is an elementary or a 
composite particle. If it is an elementary particle, with renormalizable interactions that remain perturbative until scales
of order of the Planck scale, the natural implementation of electroweak symmetry implies a supersymmetric extension
that renders the Higgs mass parameter insensitive  to the ultraviolet physics~\cite{Martin:1997ns}. Due to the 
top-quark contribution to the loop-induced Higgs couplings, any modification of  the Higgs coupling to top-quarks~\cite{Gunion:1984yn} 
will induce a similar modification of  its coupling to gluons 
as measured in terms of their SM values. These two contributions may be rendered independent in the presence of light superpartners 
of the top-quark (stops) which could contribute in a relevant way to the loop-induced Higgs couplings. Based on this observation, 
an analysis of the possible enhancement of the Higgs couplings to top-quarks within supersymmetric models was recently
presented in Refs.~\cite{Badziak:2016exn,Badziak:2016tzl}. 

Alternatively, a natural electroweak symmetry breaking may be achieved by assuming that the Higgs is a composite
particle~\cite{Terazawa:1976xx,Terazawa:1979pj,Kaplan:1983fs, Kaplan:1983sm}. There have been renewed interests in composite Higgs models, following the works in Ref.~\cite{ArkaniHamed:2001nc,ArkaniHamed:2002qx,ArkaniHamed:2002qy,Contino:2003ve,Agashe:2004rs}, and their interpretations as duals of models of gauge-Higgs unification in warped extra dimensions~\cite{Hosotani:1988bm}. In these models, the Higgs appears as a pseudo-Golstone boson and the insensitivity to the ultraviolet
scale is ensured by its composite nature, as manifest by its gauge origin in the gauge-Higgs unification picture. 
The pseudo-Goldstone nature of the Higgs scalar stems from the spontaneous breakdown of a global symmetry group that 
includes the weak interaction group as a subgroup of it. One of the simplest and most attractive realization is when the global
symmetry group is $SO(5)$~\cite{Agashe:2004rs}, which breaks spontaneously into $SO(4)$, that contains both the gauge group $SU(2)_L$, as well 
as the custodial group $SU(2)_R$. The four Nambu-Goldstone bosons associated with the breaking of the global group are 
identified with the four components of the Higgs doublet.   The properties of the Higgs boson are determined by explicit 
$SO(5)$ symmetry breaking terms associated with the Yukawa coupling of the third generation quarks,  which depend 
strongly on the representation of $SO(5)$ employed in the fermion sector.

The Higgs couplings in the minimal $SO(5)/SO(4)$ model have been previously studied in the literature~\cite{Falkowski:2007hz,Low:2009di,Low:2010mr,Montull:2013mla,Carena:2014ria}, and it is known that the simplest representation choices lead to a suppression of both the third generation quark and gluon Higgs couplings with respect to the SM ones. In this article, we provide an analytical study of the pattern of the top, bottom and gluon couplings with the Higgs within this minimal model, for different choices
of the representations in which the top quark is included.  These three couplings are of central importance to the Higgs phenomenology at the LHC. One of our goals is to provide an analytical understanding of the capabilities of this model 
to fit the future Higgs data. Moreover, we compute the Coleman-Weinberg  potential for the Higgs field that is induced by the top quark sector~\cite{Falkowski:2006vi,Medina:2007hz} and study the constraints coming from the requirement of obtaining a proper electroweak symmetry breaking (EWSB) with a Higgs mass consistent with the observed one.


The presentation of this article is as follows. In section~\ref{sec:general} we introduce a general framework for computing the relevant Higgs couplings and the Higgs potential by integrating out heavy partners of the third generation quarks in composite Higgs models. In section~\ref{sec:5and10} we focus on the minimal $SO(5)/SO(4)$ model and analyze the case of introducing composite fermions in the {\bf 5} and {\bf 10} representation of $SO(5)$.  In section~\ref{sec:14} we analyze the case of employing the {\bf 14} representation of $SO(5)$.  We
reserve section~\ref{conclusions} for our conclusions,. In the Appendices we present some technical overview and details associated with the
study. We also briefly discuss the more complicated scenario of using  {\bf 5} + {\bf 10} representations of the $SO(5)$ group in the Appendix.

\section{General analysis}
\label{sec:general}

In this section, we present a general analysis of the relation between the $t\bar{t}h$ and $ggh$ couplings, under broad assumptions that can be applied to arbitrary coset $G/H$ in composite Higgs models. We proceed by integrating out the new TeV scale strong dynamics, which results in an effective Lagrangian containing only SM particles. Effects of the strong dynamics are encoded in terms of the form factors of the SM particles in  momentum space, in analogy with the form factors of the nucleons in low-energy QCD. 
 Focusing on the third generation quarks for now,  the form factors are defined as:
\bea
\label{eq:formfactors}
&&\Pi_{t_L}\ \bar{t}_L \slashed  p\  t_L +  \Pi_{t_R}\ \bar{t}_R \slashed p \ t_R+ \Pi_{b_L}\ \bar{b}_L \slashed p\  b_L +  \Pi_{b_R} \ \bar{b}_R \slashed p \ b_R \nonumber \\
 && \qquad\qquad - (\Pi_{t_Lt_R}\ \bar{t}_L\ t_R + \Pi_{b_Lb_R}\ \bar{b}_L\ b_R + {\rm h.c.}) 
\eea
where  the form factors are the functions of $p^2$ and the proto-Yukawa couplings. 

We will also assume that  SM fermions obtain their masses from  linear mixing with the new strong sector according to the hypothesis of \emph{partial compositeness} \cite{Kaplan:1991dc}, which means in the UV, we have the mixing Lagrangian:
\beq
\label{eq:mixL}
\mL_{mix} = (\bar{q}_{L})_\alpha (y_L)^\alpha_{\ I}\mO^I_{q_L}+ \bar{t}_{R} (y^t_{R })_I \mO^I_{t_R} +  \bar{b}_{R} (y^b_{R })_I \mO^I_{b_R}
\eeq
where the operators $\mO^I_i$ from the strong sector furnish some linear representations of $G$. Note that ${\cal L}_{mix}$ must break $G$ explicitly and the proto-Yukawa couplings $y_{L,R}$ can be viewed as spurions parameterizing the effects of the explicit breaking. Then it should be clear that, after integrating out the strong dynamics, the wave function renormalizations $\Pi_{q_{L,R}}$ with $q = t, b$  are proportional to $y_{L,R}^2$, while the mass terms $\Pi_{q_Lq_R}$ are proportional to $y_L y_R$. A detailed spurion analysis could put further constraints on the form factors, as will be shown later when we discuss specific embedding of the fermionic partners.


Since we are mainly interested in composite Higgs models in which the Higgs boson is realized as a pseudo-Nambu-Goldstone boson (pNGB), we will further assume that the wave function normalization form factors can be expanded in series of $s_h^2\equiv \sin^2\frac{h}{f}$, where $f$ is Goldstone boson decay constant:
\beq
\label{eq:formexp}
\Pi_{q_L}  = \Pi_{0q_L} + s_h^2\ \Pi_{1q_L} + s_h^4\ \Pi_{2q_L}  + \cdots , \qquad \Pi_{q_R}  = \Pi_{0q_R} + s_h^2\ \Pi_{1q_R} + s_h^4 \ \Pi_{2q_R} + \cdots,
\eeq
where $q = t, b$. The expansion follows from the observation that the Higgs boson is a double under $SU(2)_L$ and that there is a shift symmetry acting on the doublet \cite{Low:2014oga}. For the form factors in front of the mass term, if the fermion is embedded in a vectorial representation $\Pi_{q_Lq_R} \sim s_{2h} \sim s_h c_h$, while for a spinorial representation it is simply $s_h$. Since the spinorial representation of the $SO(5)/SO(4)$ model is severely constrained by the precision electroweak measurements \cite{Agashe:2006at}, we will focus on the vectorial representations and its direct product:
\beq
\Pi_{q_Lq_R}  = s_h  c_h \left(\Pi_{1q_Lq_R} + s_h^2 \ \Pi_{2q_Lq_R} + \cdots \right),
\label{eq:piqlqr}
\eeq
where $\Pi_{1,2}$ are proportional to the mixing parameters $y_Ly_R$. 


To compute the Higgs couplings in models where the Higgs is a pNGB, it is important to recall that  $\left<h\right>$ is not the same as the SM Higgs vacuum expectation value of $v=246$ GeV. Instead, by matching to the $W$ boson mass in the SM one obtains \cite{Low:2014oga}
\be
v = f \sin \theta \ ,
\ee
where the misalignment angle $\theta$ is defined as $\theta = \left<h\right>/f$.
For $SO(5)/SO(4)$ coset this is explicitly demonstrated in \Eq{eq:leadingLag} of Appendix~\ref{app:CCWZ}. The misalignment angle $\theta$ is related to the fine-tuning parameter
\be
\label{eq:xidef}
\xi = \frac{v^2}{f^2} = \sin^2\theta \ ,
\ee
which is commonly employed in the literature.

The Higgs coupling to fermions can be computed by noting that the fermion masses is computed from the form factors at the zero momentum:
\beq
\label{eq:mq}
m_q =  \frac{\Pi_{q_Lq_R}(0)}{\sqrt{\Pi_{q_L}(0)}\sqrt{\Pi_{q_R}(0)}}\ ,\\
\eeq
from which we can calculate the $q\bar{q} h$ coupling strength with respect to SM as a function of the form factors:
\bea
\label{eq:cq}
c_q & \equiv& \frac{g_{q\bar{q}h}}{(g_{q\bar{q}h})_\text{SM} }= \frac{v}{m_q}  \frac{\partial m_q}{\partial \left<h\right>} =\sin\theta \frac{\partial }{\partial \theta} \log m_q  \nonumber \\
&=&  \sin\theta   \frac{\partial }{\partial \theta} \log \Pi_{q_Lq_R}  - \frac12 \sin\theta   \frac{\partial }{\partial \theta}\left( \log \Pi_{q_L} +    \log \Pi_{q_R}\right) 
\qquad {\rm at} \ \ q^2=0 \ .
\eea
It turns out that for representations considered in this work, the expansions of the form factors terminate at $\Pi_2$ and the  $q\bar{q}h$ coupling strength is given by:
 \bea
 \label{eq:cqform}
c_q
&=& \frac{\cos2\theta}{\cos\theta} +  \frac{ 2\ \Pi_{2q_Lq_R}  \sin^2\theta \cos\theta \,  }{\Pi_{1q_Lq_R} + \Pi_{2q_Lq_R} \sin^2\theta \,  } \nonumber \\
&&\qquad\qquad - \left(\frac{\Pi_{1q_L} \sin^2\theta \cos\theta  + 2\  \Pi_{2q_L}\sin^4\theta\cos\theta}{\Pi_{0q_L} + \Pi_{1q_L}  \sin^2\theta  + \Pi_{2q_L} \sin^4\theta } + L \rightarrow R\right) \qquad {\rm at} \ \ q^2=0 \ \ \ ,
\eea
where $q = t, b$. 
Note that the first term is the universal suppression factor coming from the $s_h c_h$ term in the expansion, which can be rewritten in terms of $\xi$:
\beq
 \frac{\cos2\theta}{\cos\theta}  = \frac{1 - 2 \xi}{\sqrt{1 - \xi}} \ .
\eeq

Before computing the $ggh$ coupling strength, it is worth recalling the observations made in Refs.~\cite{Azatov:2011qy,Furlan:2011uq, Montull:2013mla}, which states that, under the assumption of partial compositeness,
the determinant of the fermion mass matrix is proportional to the the mass term form factor $\Pi_{q_Lq_R}$ at the zero momentum. This is 
due to the particular form of the mass matrix:
\beq
\label{eq:mass}
\begin{split}
-\mL_m = (\bar{F}_L, \vec{\bar{\Psi}}_L) M_F(h) \left(\begin{array}{cc}
F_R\\
\vec{\Psi}_R
\end{array}\right) , \qquad M_F=  \left(\begin{array}{cc}
0 & Y^T_L(h)\\
Y_R(h) &  M_c
\end{array}
\right),\qquad 
\end{split}
\eeq
where $F$ denotes SM fermions and $Y_{L}$ ($Y_{R}$) is the mixing vector in the flavor space between the left-handed (right-handed) SM fermion $F$ and its composite  partners. Here $M_c$ is the $G$-symmetry-preserving mass matrix of the fermionic partners and does not depend on the Higgs field, because in the limit of zero mixing between SM and the composite sector, the G-symmetry is exact and all Higgs interactions must be derivatively coupled. For simplicity, we will assume that all  mixing parameters are real and have chosen a basis in the flavor space where $M_c$ is diagonal.  It is then not difficult to see:
\beq
\Det\, M_F = - Y_L^T M_c^{-1} Y_R  \ \Det\,M_c\\
\eeq
By integrating out the fermion partners using the equation of motion at the zero momentum from the Lagrangian in \Eq{eq:mass}, we  obtain:
\beq
\Pi_{F_L F_R}(0) = - Y_L^T M_c^{-1} Y_R
\eeq
which in turn implies:
\beq
\Det\, M_F = \Pi_{F_L F_R}(0) \, \Det\, M_c\\
\eeq
Note that the Higgs dependence of the determinant is fully contained in the mass form factor  $\Pi_{F_L F_R}(0)$. 

In the SM the largest contribution to the $ggh$ coupling comes from the top quark. A detailed discussion of the $ggh$ coupling is given in Appendix \ref{app:ggh}. Here we merely collect the essential results.
In the limit of infinite top mass and resonance mass,  the charge 2/3 sector  contribution to $ggh$ can be obtained:
\bea
\label{eq:cgtform}
c_g^{(2/3)} & \equiv& \frac{g^{(2/3)}_{ ggh}}{(g_{ ggh})_\text{SM}} = \sin\theta \frac{\partial }{\partial \theta} \log\Pi_{t_Lt_R} \nonumber \\
&=& \frac{\cos2\theta}{\cos\theta} +  \frac{ 2\ \Pi_{2t_Lt_R}  \sin^2\theta \cos\theta \,  }{\Pi_{1t_Lt_R} + \Pi_{2t_Lt_R} \sin^2\theta  }  \qquad {\rm at} \ \ q^2=0 \ .
\eea
For the charge -1/3 sector, the SM bottom quark contributes negligibly to the $ggh$ coupling, which need to be subtracted from the fermion mass matrix of the bottom sector.  To be specific, we have:
\bea
\label{eq:cgbform}
c_g^{(-1/3)} & \equiv &\frac{g^{(-1/3)}_{ ggh}}{(g_{ ggh)_\text{SM}}}= \sin\theta \frac{\partial }{\partial \theta} \left( \log\Pi_{b_Lb_R}  - \log m_b\right) \nonumber \\
&=& \frac{\Pi_{1b_L} \sin^2\theta \cos\theta  + 2 \ \Pi_{2b_L}\sin^4\theta\cos\theta }{\Pi_{0b_L} + \Pi_{1b_L} \sin^2\theta   + \Pi_{2b_L} \sin^4\theta } \nonumber \\
&& \qquad \qquad +  \frac{\Pi_{1b_R} \sin^2\theta \cos\theta   + 2\ \Pi_{2b_R} \sin^4\theta\cos\theta }{\Pi_{0b_R} + \Pi_{1b_R}  \sin^2\theta  +  \Pi_{2b_R}\sin^4\theta} \qquad {\rm at} \ \ q^2=0 \ .
\eea
Note that $c_g^{(-1/3)}$ starts from the linear order in $\xi$, as we have neglected the SM bottom contribution.


We study the correlation between the $ggh$ and $t\bar{t}h$ couplings by computing
\bea
\label{eq:cgct}
c_t - c_g 
&=& -\frac12 \sin\theta   \frac{\partial }{\partial \theta} \left(\log \Pi_{t_L}  +  \log \Pi_{t_R} +   \log \Pi_{b_L}  + \log \Pi_{b_R}\right)\nonumber \\
&=&-\xi \sum_{q=t,b} \left(\frac{\Pi_{1q_L}}{\Pi_{0q_L}} + \frac{\Pi_{1q_R}}{\Pi_{0q_R}} \right) + {\cal O}(\xi^2) \qquad {\rm at } \ \ q^2=0 \ .
\eea
Note that if there is no Higgs dependence for all the wave function normalization form factors, $c_t$ is exactly equal to $c_g$. We will see this limit from the specific calculations for the  different representations of $SO(5)/SO(4)$.
It turns out, at the leading order in $\xi$, $c_t-c_g$ has a strong correlation with the fermionic contribution to the Coleman-Weinberg potential of the Higgs boson, which in the Euclidean space is given by~\cite{Contino:2010rs}
\begin{equation}
\label{eq:CWpotential}
\begin{split}
V_f(h)=-2 N_c \int\frac{d^{4}Q}{(2\pi)^{4}} \left[\log\left(Q^2 \ \Pi_{t_L} \Pi_{t_R}+ |\Pi_{t_Lt_R}|^2\right)+ \log\left(Q^2\ \Pi_{	b_L}\Pi_{b_R}+ |\Pi_{b_Lb_R}|^2\right)\right] \ .
\end{split}
\end{equation}
We are only interested in the Higgs potential to the quartic order in $s_h$:
\begin{equation}
\label{eq:CWexpand}
\begin{split}
V_f(h)&\simeq-\gamma_f \ s_h^2+\beta_f \ s_h^4\\
\end{split}
\end{equation}
where 
\be
\label{eq:gammaf}
\gamma_f=  \frac{2N_c}{(4\pi)^{2}}\int_{0}^{\Lambda^2}dQ^{2}\ Q^{2}\sum_{q=t,b} \left(\frac{\Pi_{1q_L}}{\Pi_{0q_L}}+\frac{\Pi_{1q_R}}{\Pi_{0q_R}}+\frac1{Q^2}\frac{\Pi_{1q_Lq_R}^2}{\Pi_{0q_L}\Pi_{0q_R}}\right) \ ,
 \ee
while $\beta_f$ is not relevant for present discussion. In the above $\Lambda \sim 4\pi f$ is the cutoff of the composite model.  Electroweak symmetry breaking requires  
\be
\gamma_f >0 \qquad {\rm and}\qquad \beta_f >0 \ .
\ee
It turns out that, for the $SO(5)$ embedding of composite resonances studied in this work, the last contribution inside the parenthesis in Eq.~(\ref{eq:gammaf}) is numerically subleading to the first two terms, whose value at $q^2=0$ dictates $c_t-c_g$, as can be seen in Eq.~(\ref{eq:cgct}). As a result, if the integral, Eq.~(\ref{eq:gammaf}), would receive its dominant contribution from the
infrared regime, there would be a strong preference for $c_t-c_g>0$ in order to trigger EWSB. The interrelation between $c_t - c_g$ and
$\gamma_f$  will be studied in details when we consider embeddings of the composite resonances in $\bold{5}, \bold{10}$ and $\bold{14}$ of $SO(5)$.


Notice that $\gamma_f$ is quadratically divergent.   Specifically, from Eq.~(\ref{eq:formfactors}) it is clear that asymptotically in the Euclidean space,
\be
\lim_{Q^2\to \infty} \Pi_{q_{L/R}}(Q^2) \sim 1 \ , \qquad\lim_{Q^2\to \infty}  \Pi_{q_Lq_R}(Q^2) \sim  \frac1{Q^2}\ , \qquad q = t, b \ .
\ee
Under the expansion assumed in Eqs.~(\ref{eq:formexp}) and (\ref{eq:piqlqr}) we see that
\be
\lim_{Q^2\to \infty} \Pi_{0q_{L/R}}(Q^2) \sim 1 \ ,\qquad \lim_{Q^2\to \infty} \Pi_{1q_{L/R}}(Q^2) \sim \lim_{Q^2\to \infty}  \Pi_{1q_Lq_R}(Q^2) \sim \frac1{{Q^2}} \ .
\ee
These considerations suggest the quadratic divergence in $\gamma_f$ resides only in the first two terms in Eq.~(\ref{eq:gammaf}), while the last term is finite. 

 In a viable model of natural EWSB, such quadratic divergences must cancel in the Higgs potential.  The cancellation of quadratic divergent contributions to $\gamma_f$ makes the infrared contribution to Eq.~(\ref{eq:gammaf}) more relevant and therefore the correlation between $\gamma_f$ and $c_t - c_g$  stronger.  
In what follows we will always impose the cancellation of quadratic divergences in $\gamma_f$.

\section{5 and 10 of $SO(5)$}
\label{sec:5and10}

In this section, we study  the cases where the composite resonances are embedded in the $\mathbf{5}$ and $\mathbf{10}$ of $SO(5)$ and mix with the elementary fermions according to Eq.~(\ref{eq:mixL}). We will see that in neither case can the $t\bar{t}h$ coupling be enhanced over the SM expectation. However, before we begin, it is useful to set up some notation in the CCWZ formalism \cite{Coleman:1969sm,Callan:1969sn}, which will be used heavily in this work. A brief overview of CCWZ can be found in Appendix \ref{app:CCWZ}.

The main objects of consideration are the Goldstone matrix $U$ and the Goldstone gauge field $E_\mu$, defined in Eqs.~(\ref{eq:Umatrix}) and (\ref{eq:dandE}), respectively. The matrix $U$  transforms under the non-linearly realized $SO(5)$ as:
 \be
 U \to g \ U \ h^\dagger(x) \ ,
 \ee
 where $g\in SO(5), h(x) \in SO(4)$. Therefore, formally speaking, we can view the $U$ matrix as carrying an $SO(5)$ index on the left and an $SO(4)$ index on the right:
 \be
 U^{I}_{\ i} \to g^I_{\ J} \ U^J_{\ j} \ h^{\dagger j}_{\ \ i}(x)\ ,
 \ee
 where we use the capital Roman letters $I, J$ to denote the $SO(5)$ indices and the lower case Roman letters $i, j$ to denote the $SO(4)$ indices. In addition, we choose a basis such that the unbroken $SO(4)$ generators reside in the upper $4\times 4$ block of the $SO(5)$ generators.  In this basis $U^I_{\ i}$, $i=1,\cdots, 4$, can be viewed as an $SO(5)$ vector and an $SO(4)$ vector, while $U^I_{\ 5}$ transforms like an $SO(5)$ vector and an $SO(4)$ singlet. It will be useful to define $\Sigma^I$ such that
 \beq
 \label{eq:SigmaIdef}
\Sigma^I = U^I_{\ 5}  = \left(
0,0,0,s_h,c_h \right)^T, \qquad \Sigma^\dagger \Sigma = 1
\eeq 
where we have defined $s_{h} = \sin h/f, c_{h} = \cos h/f$ and evaluated the Goldstone matrix in the unitary gauge. $\Sigma$ will play a major role in building $SO(5)$ invariants. 




\subsection{5 of $SO(5)$}

We first discuss the case of embedding  the composite resonances in the  $\mathbf{5}$ of $SO(5)$ in the top sector. 
Notice that we assume $SO(5)$ is {\em explicitly} broken by the mixing of the composite resonances with the elementary fermions. 
The composite resonances transform as $\mathbf{4}$ ($\Psi_{\mathbf{4}}$) and $\mathbf{1}$ ($\Psi_{\mathbf{1}}$)
under $SO(4)$ transformations. The $SO(4)$ is unbroken in the strong sector, but it is also explicitly broken by the elementary-composite mixing. In other words, the mixing explicitly break $SO(5)$ to SM group $SU(2)_L\times U(1)_Y$. In addition, $\mathbf{4}$ decomposes into two $SU(2)_L$ doublets of hypercharge $Y = T^{3R} + X$, which are denoted as  $q_T = (T, B)^T_{1/6}$ and $q_{\mathcal{X}} = (\mathcal{X},\mathcal{T})^T_{7/6}$, where $X = 2/3$ and the subscriptsdenote the hypercharge values. More explicitly, 
\beq
\label{eq:fourplet}
\Psi_{\mathbf{4}} = \frac{1}{\sqrt{2}}\left(
\begin{array}{c}
i B - i \mathcal{X} \\
 B + \mX \\
i T + i \mT \\
 -T + \mT 
\end{array}
\right)\ ,\quad \qquad \Psi_{\mathbf{1}} = \tilde{T}.
\eeq
The Lagrangian involving the composite fermions is then: 
\bea
\label{eq:Lag55}
\mathcal{L}^{M5} &=&  i \bar{\Psi}_{\mathbf{4}} (\slashed{D} + i \slashed{E}) \Psi_{\mathbf{4}} - M_4 \bar{\Psi}_{\mathbf{4}} \Psi_{\mathbf{4}} +i \bar{\Psi}_\mathbf{1} \slashed{D} \Psi_\mathbf{1} - M_1 \bar{\Psi}_\mathbf{1} \Psi_\mathbf{1} \nonumber \\
&& \quad+\left[   c_4 y_L  f (\bar{q}^5_L)_I U^I_{\ i}(\Psi_{\mathbf{4}})^i_R +  \, a_4 y_R  f (\bar{t}^5_R)_I U^I_{\  i}(\Psi_{\mathbf{4}})^i_L  + h.c.\right]   \nonumber \\
%
&& \quad+\left[  c_1 y_L f (\bar{q}^5_L)_I U^I_{\ 5} (\Psi_\mathbf{1})_{R} +  \, a_1 y_R f (\bar{t}^5_R)_I U^I_{\ 5} (\Psi_\mathbf{1})_{L} + h.c.\right]   \ ,
\eea
where $I=1,\cdots, 5$ and $i=1,\cdots, 4$. In the above the first line contains the fermion kinetic and Dirac mass terms, while the second and the third lines contain the mixing terms for $ \Psi_{\mathbf{4}}$ and $ \Psi_{\mathbf{1}}$, respectively, with the elementary fermions $q_L$ and $t_R$, which are the explicit realization of the partial compositeness assumption in Eq.~(\ref{eq:mixL}). In particular, we have \lq\lq uplifted" the elementary fermions, which only carry $SU(2)_L\times U(1)_Y$ quantum number, to the $SO(5)$ space:
\be
\label{eq:qL5}
q_L^5 = \frac{1}{\sqrt{2}} \left(
\begin{array}{c}
i b_L \\
b_L\\
i t_L \\
- t_L \\
0
\end{array} \right) \equiv  t_L P_{t_L} + b_L P_{b_L}\ , \qquad  t_R^5 = \left(
\begin{array}{c}
0 \\
0\\
0 \\
0 \\
t_R
\end{array} \right) \equiv t_R P_{t_R} \ ,   
\ee
where we have defined the following projection operators:
\beq
\label{eq:projection5}
\begin{split}
(P_{t_L})^I = \frac{1}{\sqrt{2}} \left(
\begin{array}{c}
0 \\
0\\
i  \\
-1 \\
0
\end{array} \right), \quad 
(P_{b_L})^I = \frac{1}{\sqrt{2}} \left(
\begin{array}{c}
i \\
1\\
0  \\
0 \\
0
\end{array} \right), \quad 
(P_{t_R})^I = \left(
\begin{array}{c}
0 \\
0\\
0  \\
0 \\
1
\end{array} \right)   .
\end{split}
\eeq
In Eq.~(\ref{eq:Lag55}) the $c_i$ and $a_i$, $i=1, 4$, are dimensionless parameters that are of order unity.  Since we are not going to discuss the CP-violating effects in this paper, we assume these parameters are real.

The projection operators in Eq.~(\ref{eq:projection5}) can be viewed as spurions carrying an $SO(5)$ index:
\beq
\label{eq:projection}
(P_{t_L})^I \rightarrow g^I_{\ J} (P_{t_L})^J, \qquad (P_{t_L}^\dagger)_I \rightarrow g_I^{* J} (P_{t_L}^\dagger)_J.
\eeq
These properties allow one to construct invariants which formally respect the full $SO(5)$ symmetry of the strong sector. The elementary fermions have the following $U(1)^3_{el}$ global symmetry associated with them:
\beq
t_{L,R} \rightarrow e^{i\alpha_{L,R}} t_{L,R}, \quad P_{t_{L,R}} \rightarrow e^{-i\alpha_{L,R}} P_{t_{L,R}},\quad b_{L} \rightarrow e^{i\beta_{L}} b_{L}, \quad P_{b_{L}} \rightarrow e^{-i\beta_{L}} P_{b_{L}} .
\eeq
so that the Lagrangian for the composite Higgs has a large global symmetry $\mathcal{G} = SO(5)\times U(1)_X \times U(1)^3_{el}$ where $SO(5)$ is non-linear realized.\footnote{The $U(1)_X$ is required to give the correct hypercharges \cite{Agashe:2004rs}. The projection operators in Eq.~(\ref{eq:projection}) also transform under  $U(1)_X$.} When we integrate out the composite resonances, the resulting effective Lagrangian preserves this large symmetry $\mG$ and, as a consequence, can be constrained by performing a spurion analysis. More specifically, the effective Lagrangian can be constructed out of the following invariants: 
\beq
\label{eq:invtop5}
 P^\dagger_{t_L}\Sigma \Sigma^\dagger P_{t_L} =\frac{s_h^2}{2} ,\quad  P^\dagger_{b_L}\Sigma \Sigma^\dagger P_{b_L} =0, \quad  P^\dagger_{t_R}\Sigma \Sigma^\dagger P_{t_R} = c_h^2 , \quad P^\dagger_{t_L}\Sigma \Sigma^\dagger P_{t_R} = - \frac{s_h c_h}{\sqrt{2}}  \ ,
\eeq
where $\Sigma$ is defined in Eq.~(\ref{eq:SigmaIdef}).
This argument implies the wave function  form factors  in Eq.~(\ref{eq:formfactors}), $\Pi_{t_{L,R}}$, are invariant under $\mG$, while   $\Pi_{t_Lt_R} $  is built out of $P_{t_L}^\dagger$ and $P_{t_R}$ with similar transformation properties under ${\cal G}$.  In particular, we see $\Pi_{t_L}$ can only contain a constant term and a term proportional to $s_h^2$, while there is no dependence on the Goldston bosons (i.e. the Higgs boson) in $\Pi_{b_L}$.


The mass eigenstates before EWSB can be obtained by rotating the left-handed fields and right-handed fields with angles $\theta_{L,R}$:
\beq
\tan\theta_L = \frac{c_4 y_Lf}{M_4}, \qquad \tan\theta_R =  \frac{a_1y_R f}{M_1}
\eeq
The mixing matrices will obtain corrections  after the Higgs receives a VEV upon EWSB.
 It is straightforward to obtain the full  mass matrices by plugging \Eq{eq:fourplet}  and the expression of the Goldstone matrix $U = e^{i \Pi}$  in \Eq{eq:Uuni} into the Lagrangian in \Eq{eq:Lag55}. The result for the charge-2/3 states reads:
\beq
\begin{split}
- \Lag_{M_{2/3}} & = \left(\bar{t}_L,  \bar{T}_L,\bar{\mT}_{L},\bar{\tilde{T}}_L \right)
 M_{2/3}
\left(\begin{array}{c}
t_R \\
 T_R \\
 \mT_{R}\\
  \tilde{T}_R 
\end{array}\right), \qquad  M_{2/3} = \left(\begin{array}{cccc}
0 & Y_L^T \\
Y_R & M_c
\end{array}\right) , \\
\end{split}
\eeq
where the mixing vectors $Y_L, Y_R$ and the composite mass matrix $M_c$ are:
\beq
\label{eq:mixvec}
\begin{split}
Y_L&=  -y_L f \left(\begin{array}{c}
 c_4 \frac{1 +  \cos\theta}{2}  \\
  c_4 \frac{1 -   \cos\theta}{2} \\
c_1 \frac{\sin\theta }{\sqrt{2}} \\
\end{array}\right), \quad Y_R =  y_R f \left(\begin{array}{c}
  a_4\frac{\sin\theta}{\sqrt{2}}  \\
-a_4 \frac{\sin\theta}{\sqrt{2}}    \\
 - a_1\cos\theta \\
\end{array}\right), \quad M_c = \text{diag}(M_4, M_4, M_1).\\
\end{split}
\eeq
Recall $\theta$ is the vacuum misalignment angle defined as  $\theta = {\left<h\right>}/f$.
 The dependence on the $\sin\theta, \cos\theta$ in $Y_{L,R}$ can be understood by restoring $h$ to its full $SU(2)_L$ doublet $H$.
 
To determine the effects of the composite resonances on the Higgs couplings and the cancellation of quadratic divergences, we  need to compute:
\bea
\label{eq:det}
\Det M_{2/3} &=& - Y_L^T M_c^{-1} Y_R^* \ \Det M_c\nonumber \\
 &=&  -\frac{y_L y_R f^2}{\sqrt{2}} \sin\theta \cos\theta\left(  \frac{ c_4 a_4}{M_4}-  \frac{ c_1 a_1}{M_1}\right) M_4^2 M_1\ ,  \\
\Tr[ M_{2/3}^\dagger M_{2/3}] &=& 2 M_4^2 + M_1^2 + c_4^2 y_L^2 f^2 + a_1^2y_R^2 f^2 \nonumber \\
&&\qquad\qquad + \left[\frac{c_1^2 - c_4^2}{2}y_L^2 f^2 + (a_4^2 - a_1^2)y_R^2 f^2 \right]\sin^2\theta \ .
\eea
We can see that in the limit $c_4 = c_1, a_1 = a_4, M_1 = M_4$, the mass determinant is zero and the top quark remains massless, since the full $SO(5)$ symmetry is unbroken.
On the other hand, in the case of $c_1^2 = c_4^2, a_1^2 = a_4^2$, there is no Higgs dependence in the trace of $M_{2/3}^\dagger M_{2/3}$, which means that the quadratic divergence is cancelled in this limit. 

The $ggh$ coupling is obtained by plugging in  \Eq{eq:det} into  \Eq{eq:cgt}:
\bea
\label{eq:cgt5}
c_g &=&
\frac{\cos2\theta}{\cos\theta} =\frac{1 - 2 \xi}{\sqrt{1 - \xi}} \nonumber \\
&\equiv&    1 + \Delta_g \xi  + {\cal O}(\xi^2)\ , \qquad \qquad \Delta_g = -\frac32\ ,
\eea
where $\xi=\sin^2\theta=v^2/f^2$. Note that there is no composite mass dependence in the $ggh$ coupling \cite{Low:2009di,Low:2010mr}. This can
be understood from the fact that there is only one $\mG$-invariant  that can be constructed out of $P_{t_L}, P_{t_R}$ with Higgs dependence:
\beq
 P^\dagger_{t_L}\Sigma \Sigma^\dagger P_{t_R} = - \frac{s_h c_h}{\sqrt{2}}  
\eeq
The determinant of the mass matrix (i.e. the form factor $\Pi_{t_Lt_R}(0)$) is then proportional to $P^\dagger_{t_L}\Sigma \Sigma^\dagger P_{t_R}$ and the dependence on the mass scales can only enter through the proportionality constant, which drops out in Eq.~(\ref{eq:cgt5}).

On the other hand, to compute $c_t$ we need to compute the form factors defined in \Eq{eq:formfactors}, which can be done by following the procedure of integrating out the composite resonances outlined in Appendix \ref{app:effLag}. In terms of the expansion defined in  Eqs.~(\ref{eq:formexp}) and (\ref{eq:piqlqr}), the  form factors are
\beq
\label{eq:form5Euclidean}
\begin{split}
\Pi_{0 t_L} (p^2) &= \Pi_{0 b_L} (p^2) = 1 - \frac{c_4^2 y_L^2 f^2}{p^2 - M_4^2}, \qquad \Pi_{1 t_L} (p^2) = \frac12 y_L^2 f^2\left(\frac{c_4^2}{p^2 - M_4^2} -  \frac{c_1^2 }{p^2 - M_{1}^2} \right) \ ,  \\
\Pi_{0 t_R} (p^2) &= 1 - \frac{a_1^2 y_R^2 f^2}{p^2 - M_{1}^2}, \qquad \Pi_{1 t_R} (p^2) = y_R^2 f^2 \left(- \frac{a_4^2}{p^2 - M_4^2} +  \frac{a_1^2}{p^2 - M_1^2} \right) \ , \\
\Pi_{1t_L t_R} (p^2) &= \frac{1}{\sqrt{2}}  y_L y_R f^2  \left(\frac{ c_4 a_4 M_4 }{p^2 - M_4^2} -  \frac{ c_1 a_1   M_1 }{p^2 - M_1^2}   \right)\ ,  
\end{split}
\eeq
while all other terms vanish.
A few comments are in order. First we see the new strong sector contributions to the wave function normalization $\Pi_{t_{L}}$ and $\Pi_{t_{R}}$ are proportional to $y_{L}^2$ and $y_{R}^2$, respectively, and that to the mass term $\Pi_{t_{L}t_R}$ are proportional to $y_L y_R$. In the case where there is no $SO(5)$ breaking effects, i.e. $c_4 = c_1, a_4 = a_1, M_4 = M_1$, all the form factors vanish except the kinetic terms of the SM fermions, which are elementary fermions. On the other hand, in the limit 
\be
\label{eq:cancel}
c_4^2 = c_1^2, \qquad \qquad a_4^2 = a_1^2 \ ,
\ee 
the form factors $\Pi_{1t_{L,R}}$ vanish asymptotically, which signals the quadratic divergences in the Higgs mass cancel; see the discussion at the end of Sect.~\ref{sec:general}.  We will assume this is always the case so that the top quadratic divergence in the Higgs mass is cancelled.

To compute the top mass and the $t\bar{t}h$ coupling, we first evaluate the form factors at the zero momentum:
\beq
\label{eq:form0}
\begin{split} 
\Pi_{0t_L}(0) &= \frac{1}{\cos^2\theta_L}, \quad \Pi_{1t_L} (0)=  - \frac12 \tan^2{\theta_L} \left( 1 - \frac{1}{r_1^2} \right), \quad 
\Pi_{0t_R}(0) = \frac{1}{\cos^2\theta_R}, \\
 \Pi_{1t_R} (0) &=  -\tan^2{\theta_R} \left( 1 - r_1^2 \right), \quad \Pi_{1t_Lt_R}(0) = \frac{1}{\sqrt{2}} \frac{c_1}{c_4}M_4 \left(1 - r_1\right) \tan\theta_L \tan\theta_R  \ ,
\end{split}
\eeq
where we have defined:
\beq
r_1 = \frac{c_4 a_4}{c_1 a_1}\frac{M_1}{M_4} = \pm \frac{M_1}{M_4}\ ,
\eeq
given our choice of cancellation of quadratic divergences in Eq.~(\ref{eq:cancel}). In addition,  we set  $c_1/c_4=+1$ in the form factor $\Pi_{1t_Lt_R}(0)$, because the sign of the fermion mass term is not physical. 
Now using Eqs.~(\ref{eq:mq}) and (\ref{eq:cq}) we obtain $m_t$ and $t\bar{t}h$ coupling:
\bea
\label{eq:mtyt5}
m_t &=&  \frac{1}{\sqrt{2}}\ M_4 \left(1 - r_1\right)\ \sin\theta \cos\theta \sin\theta_L \sin\theta_R  + \cdots \ , \\
c_t & =&\frac{\cos2\theta}{\cos\theta}  - \sin^2\theta \cos\theta \left( \frac{ \Pi_{1t_L}(0) }{\Pi_{t_L}(0)} +  \frac{ \Pi_{1t_R} (0)}{\Pi_{t_R}(0)} \right)\nonumber \\
 &\equiv& 1 + \Delta_t \xi + \cdots  \ ,
\label{eq:ct5}
\eea
where we have neglected terms that are higher order in $\xi=\sin^2\theta$  in the top mass. Note that to reproduce the observed top mass for   $M_4 \sim 1\ \TeV$ and $\xi \sim 0.1$, we  need the mixing parameters $\sin\theta_L\sin\theta_R \sim {\cal O}(1)$.  

\begin{figure}[!t]
\begin{center}
\includegraphics[width=0.5\textwidth]{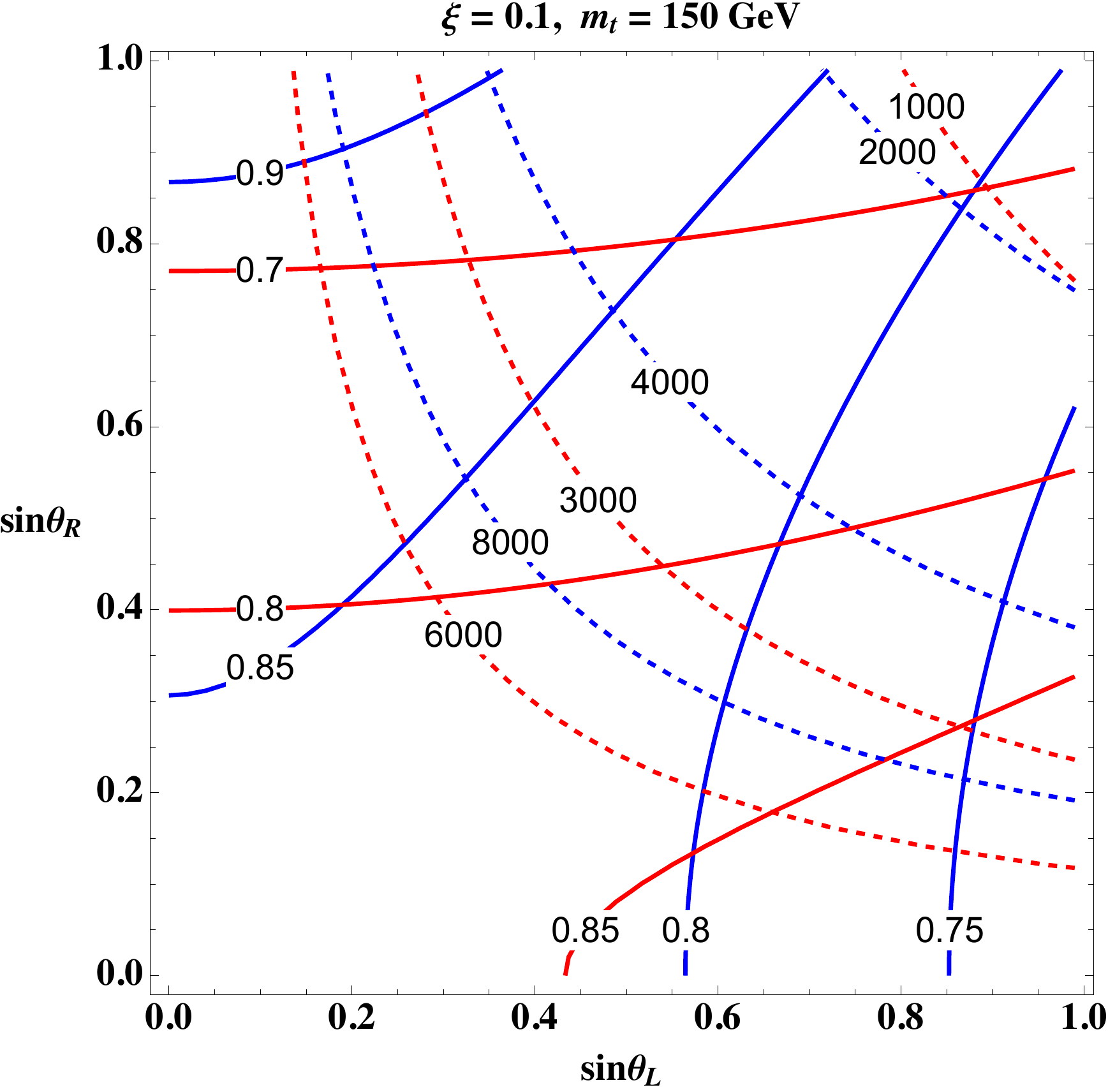}
\end{center}
\caption{Contour plots of $t\bar{t}h$ coupling strength $c_t $ (in solid lines) and the mass scale $|M_4|$ in GeV (in dashed lines) with $\xi = 0.1, m_t = 150\GeV$ for $r_1 = 0.5$ (in blue) and $r_1 = 2$ (in red). The mass scale $M_4$ is determined by the full formulae of $m_t$ using \Eq{eq:mq} in the case of $\bold{5}$. We ignore the  Higgs potential  in this plot. }
\label{fig:ct5}
\end{figure}

After substituting the form factors in \Eq{eq:form0} into \Eq{eq:ct5}, the leading modification in the 
$t\bar{t}h$ coupling $\Delta_t$ is given by
\bea
\label{eq:Dt5}
\Delta_t &=& - \frac{3}{2} + \frac12   \left(1 - \frac{1}{r_1^2} \right)\sin^2\theta_L+\left(1 - r_1^2 \right)\sin^2\theta_R  \\
             &<& \frac12 \sin^2\theta_L +  \sin^2\theta_R  -\frac32 < 0  \ .
\eea
Recall that $c_t = 1+ \Delta_t \xi$. We  see  that $\Delta_t$ is always smaller than zero, implying that the $t\bar{t}h$ coupling is always suppressed.\footnote{We also checked that in the limit $t_R$ is fully composite, even though the mass matrix in Eq.~(\ref{eq:mass}) has an non-zero first diagonal entry and $Y_R$ = 0, $t\bar{t}h$ coupling is still reduced. } 
In fact, it is possible to strengthen the bound in the above and   prove that 
\be
\label{eq:deltat12}
\Delta_t < -1/2 \ .
\ee 
To see this, we observe from Eqs.~(\ref{eq:cgt5}) and (\ref{eq:Dt5}) that 
\bea
\label{eq:DtmDg5}
\Delta_t - \Delta_g &=& \frac12   \left(1 - \frac{1}{r_1^2} \right)\sin^2\theta_L+\left(1 - r_1^2 \right)\sin^2\theta_R \nonumber \\
& \leq& \left( \frac{|\sin\theta_L|}{\sqrt{2}} - |\sin\theta_R|\right)^2 < 1 \ .
\eea
where we have used the identity $a^2+b^2 \le 2\sqrt{a^2b^2}$.
 Since $\Delta_g=-3/2$ as in Eq.~(\ref{eq:cgt5}), the bound in Eq.~(\ref{eq:deltat12}) follows.
It is also worth noting that, in the region $\theta_L \sim \theta_R$, $\Delta_t \approx \Delta_g$:
\beq
\Delta_t - \Delta_g\ \leq\ \left(\frac32-\sqrt{2}\right) \sin^2\theta_{L,R}\ \sim\ 0.1\ \sin^2\theta_{L,R}\ .
\eeq
In Fig.~\ref{fig:ct5} we plot contours of $c_t$ and $M_4$  as functions of $\theta_L$ and $\theta_R$ with $\xi = 0.1$ and  $m_t = 150$ GeV (taken as a representative value of the running top quark mass at the TeV scale), for different values of $r_1$. In the figures we always use the full formulas in Eqs.~(\ref{eq:mq}) and (\ref{eq:cq}), which captures the full dependence on $\xi$. For  $\xi=0.1$ the bound in Eq.~(\ref{eq:deltat12}) gives
\be
c_t \ \le\ 0.95 \quad {\rm for} \quad \xi = 0.1 \ ,
\ee
which is consistent with the values shown in Fig.~\ref{fig:ct5}. Notice that $M_4$ gives the overall mass scale of the top partners.


Up to now, we have not considered effects of the composite resonances in the Higgs potential. As discussed in Section~\ref{sec:general} , $c_t - c_g$ may be correlated with  the coefficient $\gamma_f$, defined in \Eq{eq:CWexpand}, in front of the quadratic term in the potential. Let's  define
\be
\label{eq:gammaff}
\gamma_f=  \frac{2N_c M_4^4}{(4\pi)^{2}}\int_{0}^{x_\Lambda}dx  \,  \mF(x)\ ,
 \ee
where  $x = Q^2/M_4^2, x_\Lambda = \Lambda^2/M_4^2$ and 
\beq
\label{eq:Fx}
\mF(x) = x \sum_{q=t,b} \left(\frac{\Pi_{1q_L}}{\Pi_{0q_L}}+\frac{\Pi_{1q_R}}{\Pi_{0q_R}}+\frac1{x}\frac{\Pi_{1q_Lq_R}^2}{\Pi_{0q_L}\Pi_{0q_R}}\right)\ .
\eeq
For the case of $\bold{5}$, we can obtain
\bea
\label{eq:mfx123}
\mF(x) &=& \frac{ r_1^2\sec^2\theta_L  \sec^2\theta_R}{(x +\sec^2\theta_L)(x +r_1^2 \sec^2\theta_R)}\left[-\left(\frac1{\xi}(c_t - c_g) +\mO(\xi)\right)\ x + \mF_0+ \mF_1(x)x\right] \ , \\
\mF_0&=& \frac12 \sin^2\theta_L \sin^2 \theta_R (1 - r_1)^2 = \frac{m_t^2}{\xi M_4^2}\left( 1+{\cal O}(\xi) \right) \ ,\\
\mF_1(x) &=&\sin^2\theta_L \sin^2 \theta_R (1 - r_1^2) \left(1 - \frac{1}{2 \,r_1^2} - \frac12 \frac{1}{x+1}\right) \ ,
\label{eq:f1in5}
\eea
where $\xi=v^2/f^2$ and $m_t$ is given in Eq.~(\ref{eq:mtyt5}). Note that in our parametrization, $r_1 = -1$ is the case of the maximally symmetric composite Higgs considered in Ref.~\cite{Csaki:2017cep}. In this limit, all terms in $\mF(x)$ vanish except $\mF_0$, which is coming from the mass form factor square term in \Eq{eq:Fx}. Note that in this special limit, the dependence on the Higgs field of the wave function normalization  form factors disappear, which implies  $c_t  = c_g$ exactly according to \Eq{eq:cgct}.
\begin{figure}[!t]
\begin{center}
\includegraphics[width=0.5\textwidth]{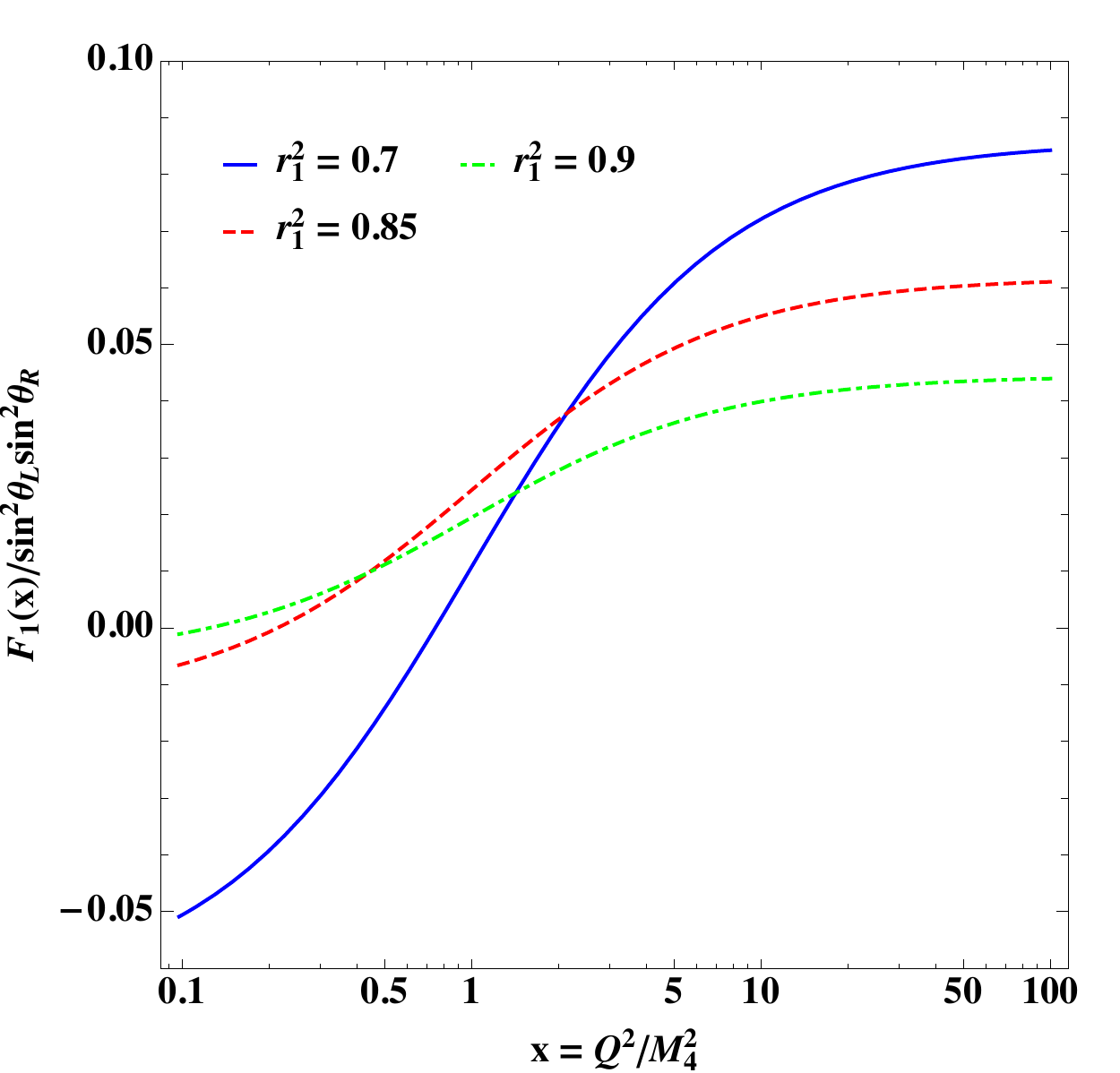}
\end{center}
\caption{ Plots for $ \mF_1(x) /\sin^2\theta_L\sin^2\theta_R$ in  the case of $\bold{5}$ for different values of $r_1^2$.}
\label{fig:DtmDgxi55}
\end{figure} 

\begin{figure}[!t]
\begin{center}
\includegraphics[width=0.45\textwidth]{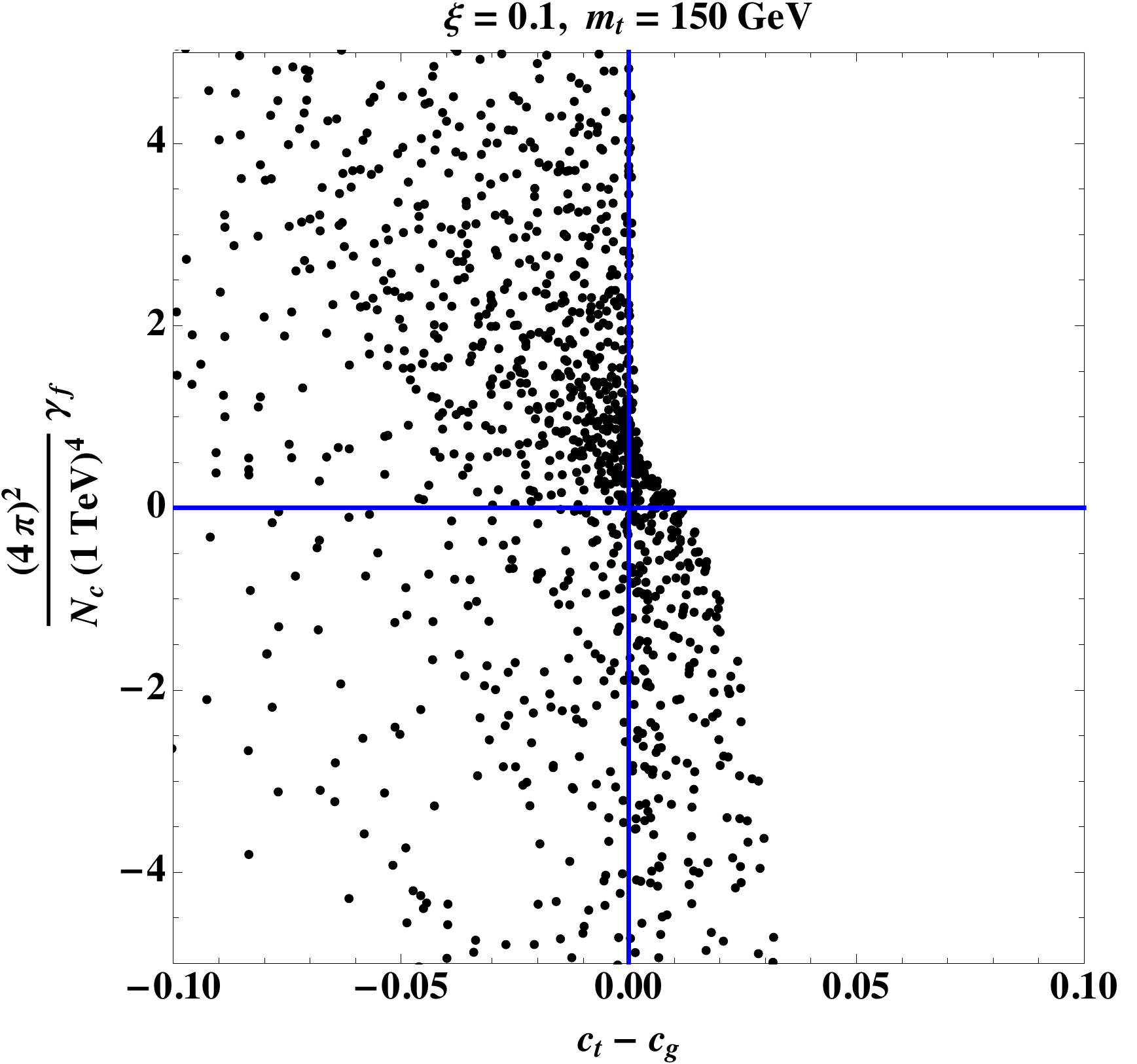}
\includegraphics[width=0.45\textwidth]{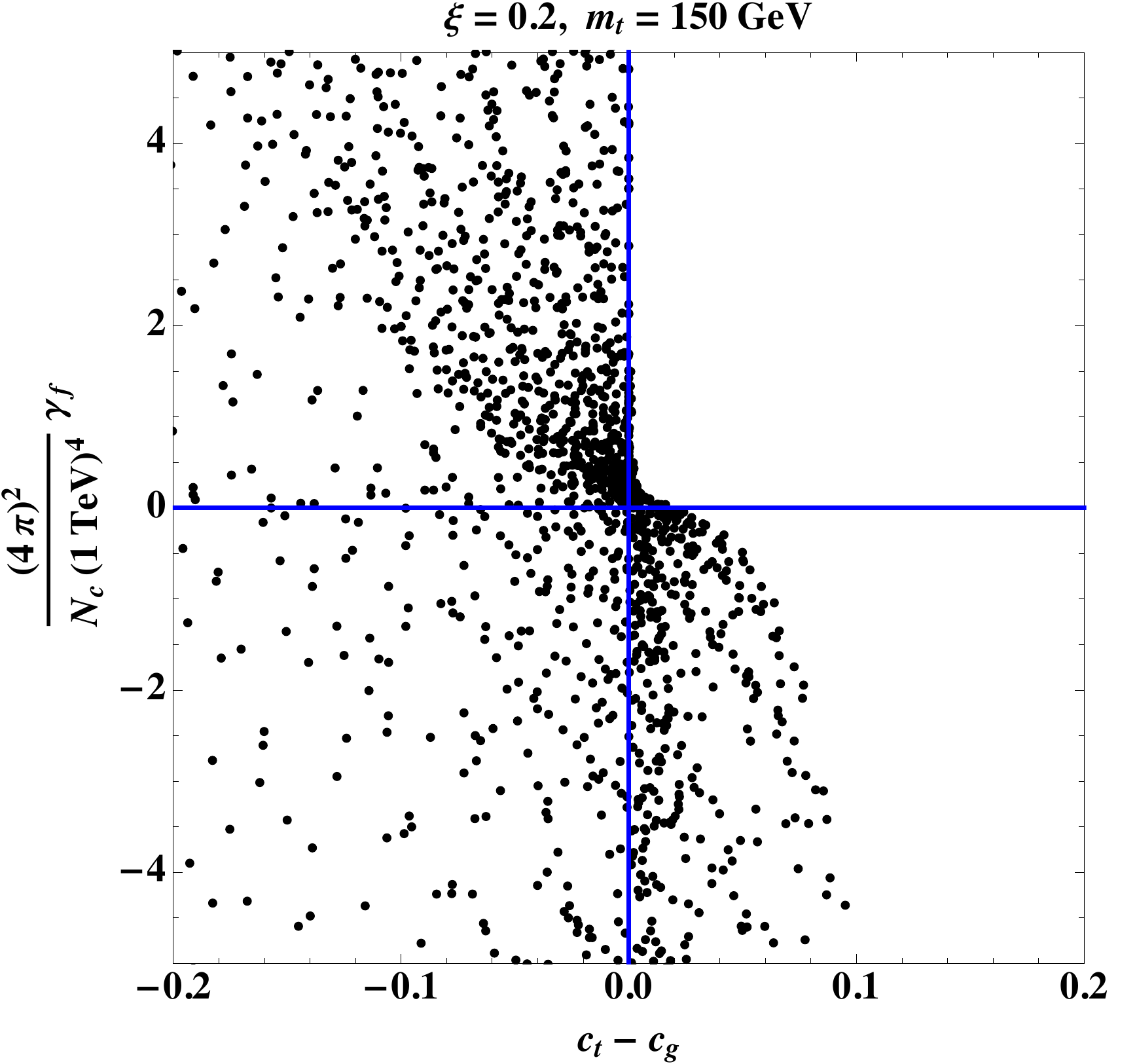}
\end{center}
\caption{ Scattering plots in the case of $\bold{5}$ for $\gamma_f$ versus $c_t - c_g$ for $\xi = 0.1$ (left panel) and $\xi = 0.2$ (right panel). We  show $\gamma_f$ in  unit of $\frac{N_c}{16\pi^2}  (1 \TeV)^4$ and fix the top mass $m_t = 150\GeV$, which is obtained by choosing the appropriate value of $M_4$. We also require that all the scales $(M_4, M_1, , c_4 y_L f, a_1 y_R f)$ are smaller than the cutoff $\Lambda = 4 \pi f$ and the lightest top partner is heavier than 500 GeV, i.e. Min$(|M_4|, \sqrt{M_1^2 + a_1^2 y_R^2 f^2}) > 500 \GeV$.}
\label{fig:gammaf5}
\end{figure}

It is worth recalling that a sufficient (although not necessary) condition for EWSB is $\mF(x) \ge 0$ through out the integration region. Among the three contributions in Eq.~(\ref{eq:mfx123}), ${\cal F}_0$ is positive-definite and constant in $x$. If one chooses $M_4\sim 1$ TeV and $\xi \sim {\cal O}(0.1)$, ${\cal F}_0 \alt {\cal O}(0.5)$ and is numerically small.  On the other hand, the first and the last terms in Eq.~(\ref{eq:mfx123}) both grows with $x$ and, therefore, should dominate the integration in $\gamma_f$. As can be seen, the first contribution is strongly correlated with $c_t-c_g$. Thus if $c_t$ is larger than $c_g$ in any significant way, one would need a sizeable  positive contribution from ${\cal F}_1(x)$ to obtain a positive $\gamma_f$ and EWSB. It turns out that  ${\cal F}_1(x)$ can be positive only in the region
\be
\frac12 \le r_1^2 \le 1 \ .
\ee
Even in this region, ${\cal F}_1(x)$ is numerically small, 
\be
{\cal F}_1(x) \le \left(\frac32 - \sqrt{2}\right) \ \sin^2\theta_L \sin^2\theta_R \sim  0.086\ \sin^2\theta_L \sin^2\theta_R \ ,
\label{eq:boundf1in5}
\ee
In Fig.~\ref{fig:DtmDgxi55} we plot ${\cal F}_1(x)$ normalized to $\sin^2\theta_L\sin^2\theta_R$ for $r_1^2 = 0.7, 0.85$ and 0.9, to demonstrate the bound in Eq.~(\ref{eq:boundf1in5}).
We conclude that the first term in Eq.~(\ref{eq:mfx123}), $-(c_t-c_g)/\xi \ x$, is the dominant contribution to $\gamma_f$ generically, and that EWSB prefers
\be
c_t\  \alt\  c_g \ .
\ee
We will see that this pattern persists considering embeddings in $\bold{10}$ and $\bold{14}$ of $SO(5)$.  In Fig.~\ref{fig:gammaf5} we present numerical scans of $\gamma_f$ versus $c_t- c_g$ for $\xi = 0.1, 0.2$, confirming  the correlation derived from the analytical understanding. In the tiny sliver of region where $c_t>c_g$ and $\gamma_f>0$, we see $c_t -c_g$ is very small, at the percent level.  Note that because the  SM gauge boson contribution to the $\gamma$ factor is  always  negative, including it will make the preference for $c_t < c_g$ even stronger.


To complete our discussion of the case of $\mathbf{5}$, we next discuss the implementation of the bottom Yukawa coupling. In this case we will introduce composite resonances to mix with $q_L$ and $b_R$, but not $t_R$.
Similar to the top partners, the bottom partners can be embedded in the $\mathbf{5}$ of $SO(5)$ but with a different $U(1)_X$ charge, $X = -1/3$. This has the effect of mixing $q_L$ with the $T^{3R} = +1/2$ doublet of the composite resonances, as opposed to the $T^{3R} = -1/2$ doublet in the case of the top partner.
The projection operators in this case   are given by:
\beq
\begin{split}
(P_{t_L})^I = \frac{1}{\sqrt{2}} \left(
\begin{array}{c}
-i \\
1\\
0 \\
0 \\
0
\end{array} \right), \qquad 
(P_{b_L})^I = \frac{1}{\sqrt{2}} \left(
\begin{array}{c}
0\\
0\\
i  \\
1 \\
0
\end{array} \right), \qquad 
(P_{b_R})^I = \left(
\begin{array}{c}
0 \\
0\\
0  \\
0 \\
1
\end{array} \right)
\end{split}
\eeq
Because of the similarity of these projection operators to their counter parts in the top sectors, all the formulas for the form factors remain the same except that the mass scales are now for the bottom partners.  The leading term for the bottom mass reads:
\bea
\label{eq:mb5}
m_b &=& \frac{1}{\sqrt{2}}\ M^{(b)}_4 \left(1 - r^{(b)}_1\right)\  \sin\theta \cos\theta  \sin\theta^{(b)}_L\sin\theta^{(b)}_R  \ .
\eea
Note that in order to reproduce the bottom mass for  $M_4^{(b)} \sim 1\ \TeV$ and $\xi \sim 0.1$, we  need $\sin\theta^{(b)}_L\sin\theta^{(b)}_R \sim 0.02$. This implies, unless we have a large hierarchy between the left-handed and the right-handed mixing parameters, the contributions to the Higgs potential from the bottom sector can be safely neglected.
 
 Now, by using Eqs.~(\ref{eq:cqform}) and (\ref{eq:cgbform}), we obtain 
\bea
\label{eq:5deltab}
\Delta_b &=& -\frac32 - \Delta_g^{(b)} \ , \\
\Delta_g^{(b)} &=& - \frac12   \left(1 - \frac{1}{(r^{(b)}_1)^2} \right) \sin^2\theta^{(b)}_L- \left(1 - (r^{(b)}_1)^2 \right)\sin^2\theta^{(b)}_R,
\eea
where we have used notations similar to those in the top sector $ c_b  = 1 + \Delta_b\xi$  and $c_g^{(b)} = \Delta_g^{(b)} \xi$.
Since we are neglecting the small SM bottom contribution to the $ggh$ coupling, the bottom partner contribution to $ggh$ coupling  in $c_g^{(b)}$ starts at the linear order in $\xi$. Notice that $\Delta_b$ in Eq.~(\ref{eq:5deltab}) has the same expression as $\Delta_t$ in Eq.~(\ref{eq:Dt5}), with all the parameters now refer to the bottom sector. Therefore,
\be 
\Delta_b < -\frac12 \ .
\ee
Because the total width of the 125 GeV Higgs boson is dominated by the partial width in the $b\bar{b}$ channel, a reduction in the bottom Yukawa could result an overall increase in the observed signal strength across a variety of channels at the LHC, without modifying the production cross-section in the $ggh$ channel.   For example, one can choose so that
\beq
\label{eq:cb5}
c_g^{(b)} \sim 0, \qquad c_b \sim \frac{1 - 2 \xi}{\sqrt{1-\xi}} \sim 1-\frac32 \xi \ ,
\eeq
which implies  $c_b \sim 0.85$ for $\xi \sim 0.1$.


\subsection{10 of $SO(5)$}

In this subsection, we embed the composite resonances in the two-index anti-symmetric representation  $\mathbf{10}$ of $SO(5)$, which can be decomposed to $(\mathbf{3},\mathbf{1})\oplus(\mathbf{1},\mathbf{3})\oplus \mathbf{4}$ under the unbroken $SO(4)$. Here we assume that the $(\mathbf{3},\mathbf{1})$ and $(\mathbf{1},\mathbf{3})$ have the same mass, which is combined into a $\mathbf{6}$ under $SO(4)$ and denoted by $\Psi^{ij}$, with $\Psi^{ij}=-\Psi^{ji}$, for $i,j = 1,\cdots, 4$.
We start from the top sector and the effective Lagrangians for the top partner fields:
\bea
\label{Lag10}
\mathcal{L}^{M6_{{10}}} & = &i \bar{\Psi}_{ij} \slashed{D} \Psi^{ji} -  \bar{\Psi}_{ij} \slashed{E}^{ji}_{\,\, kl}\Psi^{lk}  - M_6 \bar{\Psi}_{ij} \Psi^{ji} \nonumber \\
&&\qquad \qquad +\left[ c_6  y_L  f\ (\bar{q}^{10}_L)_{IJ}U^I_{\ i} U^J_{\ j}\Psi^{ji}_R +  \, a_6 y_R  f\ (\bar{t}^{10}_R)_{IJ} U^I_{\  i} U^J_{\ j} \Psi^{ji}_L  + h.c.\right]   \\
\mathcal{L}^{M4_{10}} & =& i \bar{\Psi}_i \slashed{D} \Psi^i  -  \bar{\Psi}_i   \slashed{E}^i_{\, j} \Psi^j - M_4 \bar{\Psi}_i \Psi^i \nonumber \\
&& \qquad \qquad +  \sqrt{2}\left[ c_4  y_L  f\ (\bar{q}^{10}_L)_{IJ}U^I_{\ i} U^J_{\ 5}\Psi^i_R +  \,  a_4 y_R  f \ (\bar{t}^{10}_R)_{IJ} U^I_{\  i} U^J_{\ 5} \Psi^i_L  + h.c.\right]   
\eea
where $D_\mu = \partial_\mu + i( 2/3) \, B_\mu$. 
In the limit in which the $c_4 = c_6$, $a_4=a_6$ and $M_4 = M_6$, the $SO(5)$ invariance is restored.  
We assign two upper indices to the sixplet composite fields $\Psi^{ij}$ and as a result $E_\mu$ will have two upper and two lower indices. The \lq\lq uplifting" of the elementary fermions $q_L$ and $t_R$ to the $SO(5)$ space in this case is defined by
 \beq
 \label{eq:qL10}
q_{L}^{10} = t_L P_{t_L} + b_L P_{b_L} , \qquad t_{R}^{10} =  t_R P_{t_R} \ ,
\eeq
where 
\beq
\label{eq:projection10}
\begin{split}
(P_{t_L})^{IJ}&= \frac{1}{2} \left(
\begin{array}{cccccc}
&&&&0\\
&&&&0\\
&&&& i  \\
&&&&- 1 \\
0 & 0 &- i  &  1&
\end{array} \right), \quad
(P_{b_L})^{IJ}= \frac{1}{2} \left(
\begin{array}{cccccc}
&&&&i\\
&&&&1\\
&&&& 0  \\
&&&& 0 \\
-i & -1 & 0  & 0&
\end{array} \right), \\
(P_{t_R})^{IJ}&= \frac{1}{2}  \left(
\begin{array}{cccccc}
0&-i&&& \\
i&0&&&\\
&&0&-i&  \\
&&i&0& \\
 &  &  &   & 0
\end{array} \right)\ ,
\end{split}
\eeq
which carry two $SO(5)$ upper indices and will be treated as spurions. Observe that the projection tensors are just the anti-symmetrized version of the product of the embedding vectors in the case of $\mathbf{5}$. 

Similar to the case of $\mathbf{5}$, the spurion analysis reveals only one invariant for each projection operator (keep in mind our contraction convention):
\beq
 \Sigma^T P_{q}^\dagger P_{q} \Sigma^*
\eeq
where  $q= t_{L,R}$ and $b_{L,R}$. To see this, we observe that the invariant involving the $U^I_{\ i}$ can always be reduced to the above by using the unitary constraints:
\beq
U^{I}_{\ i}U^{\dagger i}_{\ J} = \delta^I_{\ J} - \Sigma^I \Sigma^\dagger_J \ .
\eeq 
 To be specific, we have:
\beq
\label{eq:invtop10}
\begin{split}
 \Sigma^T  P_{t_L}^\dagger P_{t_L} \Sigma^*  &= \frac12 - \frac14 s_h^2,   \qquad
  \Sigma^T  P_{b_L}^\dagger P_{b_L} \Sigma^* = \frac12 - \frac12 s_h^2, \\
 \qquad \Sigma^T  P_{t_R}^\dagger P_{t_R}  \Sigma^*& = \frac14 s_h^2 \ , \qquad  \Sigma^T P_{t_L}^\dagger P_{t_R} \Sigma^* = - \frac{1}{4} s_h c_h
 \end{split}
\eeq
which implies  the expansion for the form factors in \Eq{eq:formexp} stops at $s_h^2$. 

Calculating the form factors as before, the form factors for the left-handed sector are
\beq
\begin{split}
\Pi_{0 t_L} (p^2) &= 1 -   \frac{c_4^2 y_L^2 f^2}{p^2 - M_4^2}, \qquad \Pi_{1 t_L} (p^2) = \frac12   y_L^2 f^2\left(  \frac{c_4^2 }{p^2 - M_4^2}-\frac{c_6^2 }{p^2 - M_6^2}   \right)  \ , \\
 \Pi_{0 b_L} (p^2) &= 1 -   \frac{c_4^2 y_L^2 f^2}{p^2 - M_4^2}, \qquad \Pi_{1 b_L} (p^2) = y_L^2 f^2\left( \frac{c_4^2 }{p^2 - M_4^2}   - \frac{c_6^2 }{p^2 - M_6^2}   \right)\  .
 \end{split}
\eeq
Similarly, for the right-handed sector,
\beq
\begin{split}
\Pi_{0 t_R} (p^2) &= 1 -  \frac{a_6^2 y_R^2 f^2}{p^2 - M_{6}^2}, \qquad \Pi_{1 t_R} (p^2) =   \frac12 y_R^2 f^2\left( - \frac{a_4^2}{p^2 - M_4^2} + \frac{ a_6^2 }{p^2 - M_6^2}\right) \\
 \end{split}
\eeq
and the mass term:
\beq
\Pi_{1t_L t_R} (p^2) = \frac{1}{2} y_L y_R   f^2\left(\frac{ c_6 a_6  M_6 }{p^2 - M_6^2}  -\frac{c_4 a_4  M_4 }{p^2 - M_4^2}\right)\ .   
\eeq
From the above, we  see clearly that $\Pi_{1q}$ vanishes  in the limit  $c_6=c_4$, $a_6 = a_4$ and $M_4=M_6$, and there is no
Higgs dependence in the form factors. This is expected because in this limit the full $SO(5)$ symmetry is restored and the Goldstone field can be rotated away  by redefinition of the the composite fields.

Cancellation of quadratic divergence in the top sector requires
\beq
\label{eq:solution10}
c_6^2 = c_4^2, \qquad a_6^2 = a_4^2 \ ,
\eeq
but not $M_4^2=M_6^2$, since the mass term only breaks $SO(5)$ "softly." We  assume  Eq.~(\ref{eq:solution10}).

We also define the following useful parameters
\beq
\label{eq:thetaLR}
\tan\theta_L = \frac{c_4 y_L f}{M_4}, \qquad \tan\theta_R = \frac{a_6 y_R f}{M_1}, \qquad r_6 = \frac{c_4 a_4}{c_6 a_6}\frac{M_6}{M_4}= \pm \frac{M_6}{M_4}\ .
\eeq
Now it is straightforward to obtain the top mass at the leading order in $\xi$,
\beq
\label{eq:mt14}
m_t = \frac{1}{2}M_4  \left( 1-r_6 \right) \ \sin \theta \cos \theta \sin\theta_L\sin\theta_R  \ ,
\eeq
which is the same as in the case of $\mathbf{5}$ except the $1/2$ factor in front. The modifications to the $t\bar{t}h$ and $ggh$ couplings from the top sector are then: 
\bea
 \Delta_g^{(t)}  &=&  - \frac32 + \left(\frac{1}{r_6^2} - 1\right) \sin^2\theta_L, \\
 \Delta_t &= &- \frac32 + \frac12\sin^2\theta_L \left(1 -  \frac{1}{r_6^2}\right)  +\frac12\sin^2\theta_R   \left( 1 - r_6^2 \right) \nonumber \\
            &<& -\frac32 +\frac12 (|\sin\theta_L| -|\sin\theta_R|)^2 < -1   \ ,
\label{eq:10ct11}           
\eea
where we have used  the same convention  for the $\Delta$'s as in the case of $\mathbf{5}$. We see immediately  that 1) $c_g=1+\Delta_g^{(t)}\xi$ can be either enhanced or reduced and 2) the top Yukawa coupling is not only always suppressed, but the suppression is in general  stronger than in the case of $\mathbf{5}$. On the other hand,
\bea
 \Delta_t -\Delta_g^{(t)} &=&  \frac32\sin^2\theta_L \left(1 -  \frac{1}{r_6^2}\right)  +\frac12\sin^2\theta_R   \left( 1 - r_6^2 \right)  \nonumber \\
 &<& \frac12 \left(\sqrt{3}\ |\sin\theta_L| -|\sin\theta_R|\right)^2 < \frac32 \ .
\eea

\begin{figure}[!t]
\begin{center}
\includegraphics[width=0.45\textwidth]{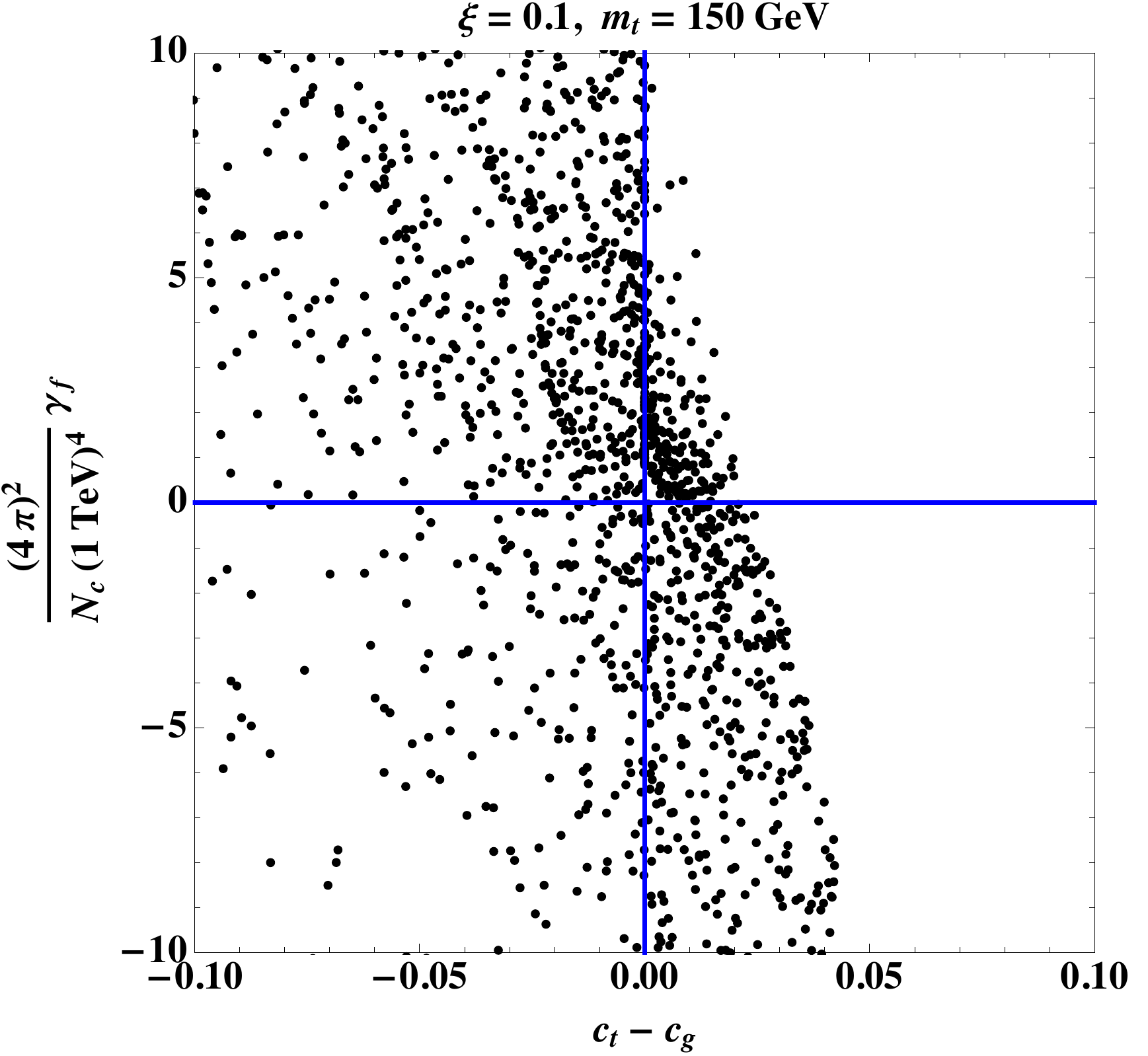}
\includegraphics[width=0.45\textwidth]{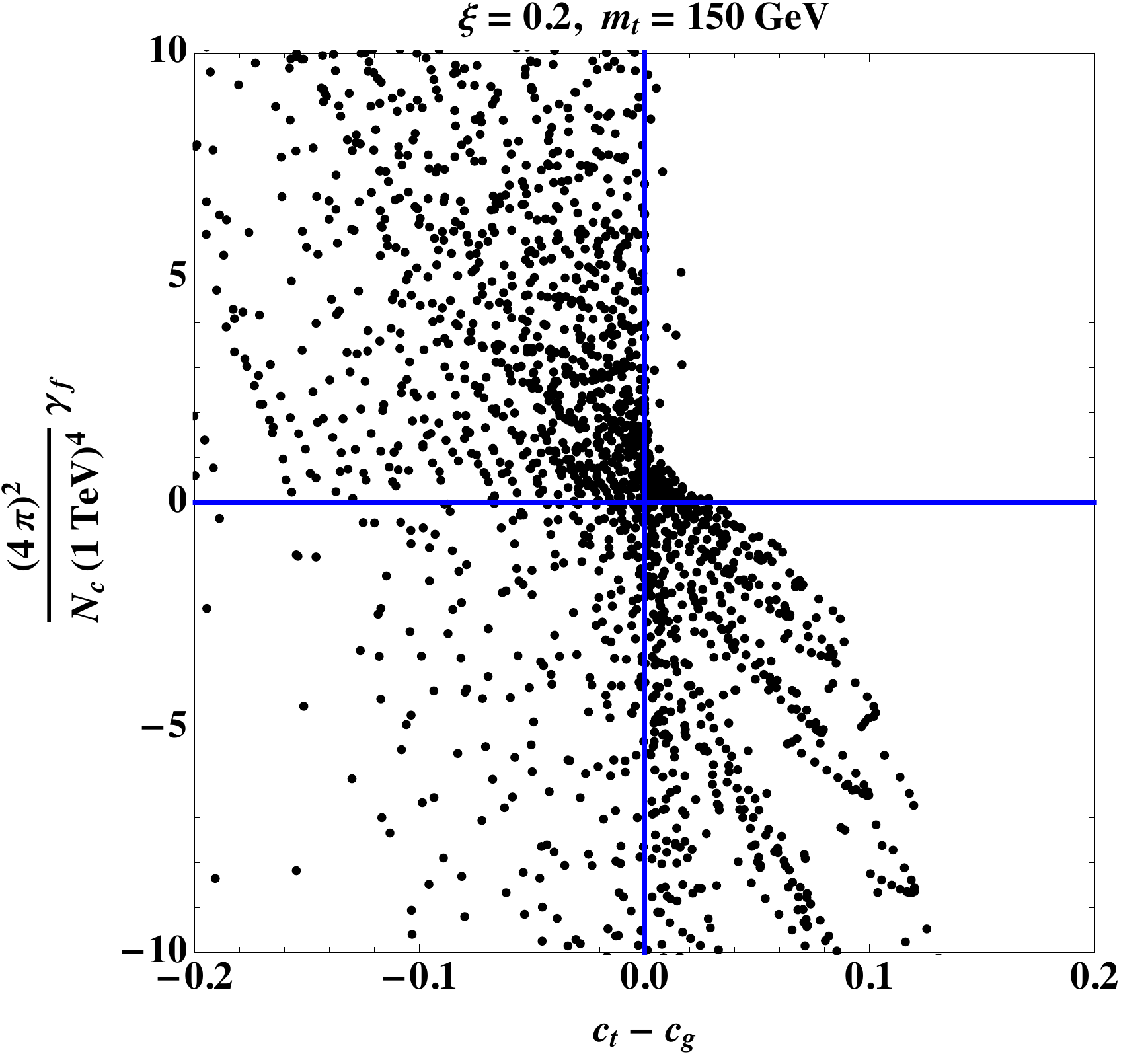}
\end{center}
\caption{ Scattering plots in the case of $\bold{10}$ for $\gamma_f$ versus $c_t - c_g$ for $\xi = 0.1$ (left panel) and $\xi = 0.2$ (right panel). We  show $\gamma_f$ in  unit of $\frac{N_c}{16\pi^2}  (1 \TeV)^4$ and fix the top mass $m_t = 150\GeV$, which is obtained by choosing the appropriate value of $M_4$. We also require that all the scales $(M_4, M_6, , c_4 y_L f, a_6 y_R f)$ are smaller than the cutoff $\Lambda = 4 \pi f$ and the lightest top partner is heavier than 500 GeV, i.e. Min$(|M_4|, \sqrt{M_6^2 + a_1^2 y_R^2 f^2}) > 500 \GeV$.}
\label{fig:gammaf10}
\end{figure}

As for the Higgs potential, we compute ${\cal F}(x)$ in Eq.~(\ref{eq:gammaff}):
\beq
\label{eq:mfx12}
\mF(x) = \frac{ r_6^2\sec^2\theta_L  \sec^2\theta_R}{(x +\sec^2\theta_L)(x +r_6^2 \sec^2\theta_R)}\left[- \left(\frac1{\xi}(c_t - c_g)  + \mO(\xi)\right)\ x + \mF_0+ \mF_1(x)x\right] \ ,
\eeq
where 
\bea
\mF_0&=& \frac14 \sin^2\theta_L \sin^2 \theta_R (1 - r_6)^2  = \frac{m_t}{\xi M_4^2}\left(1 + {\cal O}(\xi)\right)\ ,\\
\mF_1(x) &=&\frac12\sin^2\theta_L \sin^2 \theta_R (1 - r_6^2) \left[1 - \frac{3}{\,r_6^2} - \frac12 \frac{1}{x+1}+\frac52 \frac{1}{x+r_6^2}\right] \ .
\eea
Notice that ${\cal F}(x)$ has a similar structure to the case of $\bold{5}$, where ${\cal F}_0$ is related to the top mass and expected to be subdominant for $M_4\sim 1$ TeV and $\xi \sim {\cal O}(0.1)$. As for ${\cal F}_1(x)$, one can show that it is positive only in the region
\be
1 < r_6^2 < 3 \ .
\ee
In this region it is possible to demonstrate that 
\be
{\cal F}_1(x)\le 0.26 \ \sin^2\theta_L\sin^2\theta_R \ .
\ee
So again there is a strong preference for $c_t < c_g$ in order to trigger EWSB, which is confirmed in the numerical scan shown in Fig.~\ref{fig:gammaf10}.


For the bottom sector, we  again introduce composite fermions that mix with $b_R$ but not $t_R$, which is similar to the case of $\bold{5}$ and the results for the modifications of $ggh$ coupling and $hb\bar{b}$ coupling read:
\bea
\Delta_b &=& -\frac32 - \Delta_g^{(b)}  \ , \\
\Delta_g^{(b)} &=& - \sin^2\theta^{(b)}_L  \left(1 - \frac{1}{(r^{(b)}_6)^2} \right) - \sin^2\theta^{(b)}_R \left(1 - (r^{(b)}_6)^2 \right),
\eea
Again $\Delta_b$ is similar to its corresponding expression of $\Delta_t$ in Eq.~(\ref{eq:10ct11}). As a result, the same bound
\be
\Delta_b < -1 
\ee
applies. Also similar to the case of $\bold{5}$, if we assume that the mixing parameters in the bottom sector are small, it is  possible to modify the $b\bar{b}h$ coupling without chaning the $ggh$ coupling, as done in \Eq{eq:cb5}.

\section{$\bold{14}$ of $SO(5)$}
\label{sec:14}

$\mathbf{14}$ is the two-index symmetric-traceless  representation  of $SO(5)$. This scenario is distinct from the cases of $\mathbf{5}$ and $\mathbf{10}$ in that there are two non-trivial $SO(5)$ invariants one can construct in the spurion analysis. As a result, the $ggh$ coupling now has a non-trivial dependence on the mass of the composite resonance \cite{Azatov:2011qy} and, moreover, the $t\bar{t}h$ coupling can be enhanced. A qualitatively similar, but numerically more complicated, scenario of embedding the composite fermions in $\bold{5}+\bold{10}$ simultaneously is discussed in Appendix \ref{append:5+10}.



\subsection{The Top Sector}

Under the unbroken $SO(4) \simeq SU(2)_L\times SU(2)_R$, $\mathbf{14}$  can be decomposed into $\bold{9} \oplus \bold{4} \oplus \bold{1} \simeq (\bold{3},\bold{3})\oplus  (\bold{2},\bold{2})\oplus \bold{1}$. Therefore we introduce three mass scales, $M_9, M_4$ and $M_1$ for the 9-plet, 4-plet and the singlet, respectively. The uplifting of the elementary fermion to the $SO(5)$ space achieved through the following projection operators:
\be
q_{L}^{14} = t_L P_{t_L} + b_L P_{b_L} \ ,\quad
t_R^{14} 
= t_R P_{t_R} \ , 
\ee
\beq
\begin{split}
(P_{t_L})^{IJ}&= \frac{1}{2} \left(
\begin{array}{cccccc}
&&&&0\\
&&&&0\\
&&&& i  \\
&&&&- 1 \\
0 & 0 & i  & - 1&
\end{array} \right), \qquad
(P_{b_L})^{IJ}= \frac{1}{2} \left(
\begin{array}{cccccc}
&&&&i\\
&&&&1\\
&&&& 0  \\
&&&& 0 \\
i & 1 & 0  & 0&
\end{array} \right), \\
(P_{t_R})^{IJ}&= \frac{1}{2\sqrt{5}} \left(
\begin{array}{cccccc}
-1&&&&\\
&-1&&&\\
&&-1&&   \\
&&&-1&  \\
&&&&4
\end{array} \right).
\end{split}
\label{eq:projection14}
\eeq
These projection operators carry two $SO(5)$ indices, which are just the symmeterized version of the tensor product of two projection operators in the case of $\bold{5}$.

The effective Lagrangians for the top partner fields:
\bea
\label{eq:Lagtop14}
\mathcal{L}^{M9_{14}} & =& i \bar{\Psi} (\slashed{D} + i \slashed{E}) \Psi - M_9 \bar{\Psi}_{ij} \Psi^{ji} \nonumber \\
&&\qquad+\left[ c_{9}  y_L  f (\bar{q}^{14}_L)_{IJ}U^I_{\ i} U^J_{\ j}\Psi^{ji}_R +  \, a_{9} y_R  f (\bar{t}^{14}_R)_{IJ} U^I_{\  i} U^J_{\ j} \Psi^{ji}_L  + h.c.\right]   \\
\mathcal{L}^{M4_{14}} & =& i \bar{\Psi} (\slashed{D} + i \slashed{E}) \Psi - M_4 \bar{\Psi} \Psi  \nonumber \\
&& \qquad+  \sqrt{2}\left[ c_4  y_L  f (\bar{q}^{14}_L)_{IJ}U^I_{\ i} U^J_{\ 5}\Psi^i_R +  \,  a_4 y_R  f (\bar{t}^{14}_R)_{IJ} U^I_{\  i} U^I_{\ 5} \Psi^i_L  + h.c.\right]   \\
\mathcal{L}^{M1_{14}} &=& i \bar{\Psi} \slashed{D} \Psi - M_1 \bar{\Psi} \Psi  \nonumber \\
&&\qquad+  \frac{\sqrt{5}}{2}   \left[   c_1 y_L f (\bar{q}^{14}_L)_{IJ} U^I_{\ 5}  U^J_{\ 5} \Psi_{R} + a_1 \, y_R f (\bar{t}^{14}_R)_{IJ} U^I_{\ 5}  U^J_{\ 5} \Psi_{L} + h.c.\right]   
\label{eq:Lagtop143}
\eea
where $D_\mu = \partial_\mu + i 2/3 \, B_\mu$ and the numerical factors in front of the mixing terms are such that, in the limit  in which
$c_1 = c_4 = c_9$, $a_1 = a_4 = a_9$, and $M_1 = M_4 = M_9$, the full $SO(5)$ symmetry is recovered. Notice that, in the above, the 9-plet fermion $\Psi^{ij}$ is a symmetric, traceless rank-2 tensor field with $i,j=1,\cdots, 4$, which means that only the traceless part of $ (\bar{q}^{14}_L)_{IJ}U^I_{\ i} U^J_{\ j},(\bar{t}^{14}_R)_{IJ}U^I_{\ i} U^J_{\ j} $ mixing with it.  The corresponding $E_\mu$ in the kinetic term therefore has two upper and two lower indices. 


 Similar to the case of $\mathbf{5}$, the 4-plet can be decomposed as two  $SU(2)_L$ doublets which are denoted as  $q_T = (T, B)_{1/6}$ and $q_X = (X_{5/3}, X_{2/3})_{7/6}$; see \Eq{eq:fourplet}. For the 9-plet, it can be decomposed as three degenerate $SU(2)_L$ triplets ($I_L = 1$) with hypercharge $5/3, 2/3, -1/3$ respectively. From these quantum numbers we see, before EWSB, the triplet fermions and $q_X$ cannot mix with the elementary fermions, while $q_T$ and the singlet $\tilde{T}$ can mix with the $q_L$ and $t_R$, respectively, which is similar to the case of $\mathbf{5}$. The mass spectrum before EWSB is then
\beq
M_\psi = M_9,\quad M_{T} = \sqrt{M_4^2 + c_4^2 y_L^2 f^2},  \quad  M_X = M_4, \quad  M_{\tilde{T}} = \sqrt{M_1^2 + a_1^2 y_R^2 f^2}  \ ,
\eeq
where $M_{\psi},M_{T},M_X,M_{\tilde{T}}$ are the masses of the triplets, the doublet $q_T$, the doublet $q_X$ and the singlet respectively. After EWSB, the physical masses will be corrected at $\mO(\xi)$, except the exotic electric charge $(8/3, 5/3, -4/3)$ states.  We then define the following mixing parameters:
\beq
\label{eq:thetaLR14}
\tan\theta_L = \frac{c_4 y_L f}{M_4}, \qquad \tan\theta_R = \frac{a_1 y_R f}{M_1}\ ,
\eeq
which is similar to the case of $\mathbf{5}$.

As in previous cases, we first explore the independent $SO(5)$ invariants involving the Goldstone matrix and the projection operators in Eq.~(\ref{eq:projection14}).  A useful observation in this regard is the fact that an invariant involving the $U^I_{\ i}$ can be rewritten by  using the unitary constraints:
\beq
U^{I}_{\ i}U^{\dagger i}_{\ J} = \delta^I_{\ J} - \Sigma^I \Sigma^\dagger_J\ .
\eeq 
As a result, the following decomposition applies
\beq
\label{eq:reduc}
\begin{split}
&( P^\dagger_{q})_{IJ} U^I_{\  i} U^J_{\  5}U^{\dagger i}_{\ \, K} U^{\dagger 5}_{\ \ L}  (P_{q})^{KL} 
=  \Sigma^T P^\dagger_{q}  P_{q} \Sigma^* -  \Sigma^T P^\dagger_{q} \Sigma  \ \Sigma^\dagger P_{q} \Sigma^* \ . \\
\end{split}
\eeq
In the end there are precisely two and only two invariants for the form 
\beq
\begin{split}
 \Sigma^T P_{q}^\dagger P_{q} \Sigma^*, \qquad  \Sigma^T P^\dagger_{q} \Sigma  \  \Sigma^\dagger P_{q} \Sigma^*   
   \end{split}
\eeq
where here $P_q$ generally denotes $P_{t_{L,R}}, P_{b_{L,R}}$. 

An important consequence of having two different $SO(5)$ invariants is that now we have two Higgs-dependent terms in the $\Pi_{t_Lt_R}$ and, as a result, the $ggh$ coupling depends on the composite mass scales, unlike in the case of $\mathbf{5}$ and $\mathbf{10}$. To be specific, we have:
\beq
\label{eq:invtop1}
\begin{split}
 \Sigma^T P_{t_L}^\dagger P_{t_R} \Sigma^*& = - \frac{3}{4\sqrt{5}} s_h c_h,   \qquad \Sigma^T P^\dagger_{t_L} \Sigma  \  \Sigma^\dagger P_{t_R} \Sigma^*   = \left( - \frac{2 \sqrt{5}}{5} +   \frac{\sqrt{5}}{2} s_h^2\right) s_h c_h \\
\end{split}
\eeq
which implies the mass form factor now contains two different trigonometric combinations: $s_hc_h$ and $s_h^3 c_h$. In other words, $\Pi_{2q_Lq_R}$ is now non-vanishing in Eq.~(\ref{eq:piqlqr}). A similar computation for the form factors $\Pi_{t_L}$ and $\Pi_{t_R}$ shows that they contain  expansion up to  ${\cal O}(s_h^4)$, while the expansion in $\Pi_{b_L}$ stops at $s_h^2$ because $\Sigma^T P^\dagger_{b_L} \Sigma = 0$. 
In the end, the form factors in the case of $\mathbf{14}$ have the following expansions,
\beq
\begin{split}
\Pi_{t_L}  &= \Pi_{0t_L} + s_h^2 \Pi_{1t_L} + s_h^4 \Pi_{2t_L}  , \quad \Pi_{t_R}  = \Pi_{0t_R} + s_h^2 \Pi_{1t_R} + s_h^4 \Pi_{2t_R}, \\
 \Pi_{b_L}  &= \Pi_{0b_L} + s_h^2 \Pi_{1b_L} , \qquad \quad \quad \
  \Pi_{t_Lt_R}  = s_h c_h \left(\Pi_{1t_Lt_R} + s_h^2 \Pi_{2t_Lt_R} \right) .\\
\end{split}
\label{eq:14forms}
\eeq
Similar to previous cases, cancellation of quadratic divergences in the top sector requires the following  condition,
\beq
\label{eq:solution}
c_9^2 = c_4^2 = c_1^2, \quad a_9^2 = a_4^2 = a_1^2 \ ,
\eeq
 which we impose in our analysis. Furthermore, we  define
\beq
\label{eq:r1r914}
\qquad r_1 = \frac{c_4 a_4}{c_1 a_1}\frac{M_1}{M_4} = \pm \frac{M_1}{M_4} , \qquad r_9 = \frac{c_4 a_4}{c_9 a_9}\frac{M_9}{M_4}=\pm \frac{M_9}{M_4}\ .
\eeq
  Now it is straightforward to obtain the top mass from the form factors:
\beq
\label{eq:mt14}
m_t = \frac{\sqrt{5}}{2} M_4  \left( 1-r_1 \right) \ \sin \theta \cos \theta \sin\theta_L\sin\theta_R   
\eeq
where again we have neglected the higher order terms in $\xi = \sin^2\theta$. Notice the above formula is the same as the case of $\bold{5}$ except the numerical factor $\sqrt{5}/2$. One relevant difference with respect to the $\bold{5}$ and $\bold{10}$ cases is that the top-quark mass dependence on $\xi$ include  higher order terms which are proportional to $(1-r_9)$ and not to $(1-r_1)$.  This implies that for $r_1 \to 1$, the dominant dependent of the top-quark mass is not linear on the Higgs field, what makes it difficult to generate a realistic top mass and a light Higgs boson at the same time, and also causes an unacceptably large departure of the top-quark coupling to the Higgs with respect to its SM value. Hence, a phenomenologically viable model can only be obtained if the leading contribution proportional to $(1-r_1)$, \Eq{eq:mt14}, is sizable compared to  the higher order terms proportional to  $(1-r_9)$. The formulae below are derived under this assumption.

By following the same calculation as in before, we obtain the modifications to the $t\bar{t}h$  and $ggh$ couplings:
\bea
\label{eq:Deltag14}
 \Delta_g^{(t)}  &=& - 4 - \frac32\frac {  1 - 1/r_9 }{1 - 1/r_1 }   +\left(\frac{1}{r_9^2} - 1\right) \sin^2\theta_L \ , \\
 \Delta_t &=& - 4 - \frac32\frac {  1 - 1/r_9 }{1 - 1/r_1 }  + \frac54\sin^2\theta_L \left[\left(1 -  \frac{1}{ r_9^2}\right) +\left(1- \frac{1}{ r_1^2}\right)  \right] \nonumber \\
 &&\qquad\qquad\qquad  +\frac52\sin^2\theta_R   \left( 1 - r_1^2 \right)  \ ,
 \label{eq:Deltat14}
\eea
where $c_t = 1 + \Delta_t \xi$ and $c_g^{(t)} = 1 + \Delta_g^{(t)} \xi$. The superscript in $\Delta_g^{(t)}$ indicates this is the contribution  from the top sector.
Notice that, in  Eqs.~(\ref{eq:Deltag14}) and (\ref{eq:Deltat14}),  there is a fictitious pole  at $r_1 =1$, which is just a reflection of our previously stated approximation of  keeping only the leading contribution in $\xi$ to the top quark mass, \Eq{eq:mt14}.  
Again, such a fictitious pole does not arise in the case of $\bold{5}$ and $\bold{10}$ because in these two cases $\Delta_g$ and $\Delta_t$ do not have explicit dependence on the composite scales.

 
Let's analyze the $ggh$ and $t\bar{t}h$ couplings without considering the Higgs potential, for now.  The most important distinctions from the cases of $\bold{5}$ and $\bold{10}$ is that, in the current scenario, $\Delta_g$ and $\Delta_t$ can be either positive or negative. In other words, there are enough degrees of freedom in Eqs.~(\ref{eq:Deltag14}) and (\ref{eq:Deltat14}) such that $c_g$ and $c_t$ could be either enhanced or reduced. It is easiest to demonstrate this numerically. In Fig.~\ref{fig:ctcg14} we plot contours of $c_g$ and $c_t$ for the benchmark of $\xi=0.1$ and $\sin\theta_L=\sin\theta_R=0.8$ (left panel), $\sin\theta_L=0.3, \sin\theta_R=0.9$ (right panel), where it is  clear that they can be enhanced or reduced over the SM expectations for the reasonable values of $|M_4|$, which is the overall mass scale for the top partners.


 \begin{figure}[ht]
\begin{center}
\includegraphics[width=0.48\textwidth]{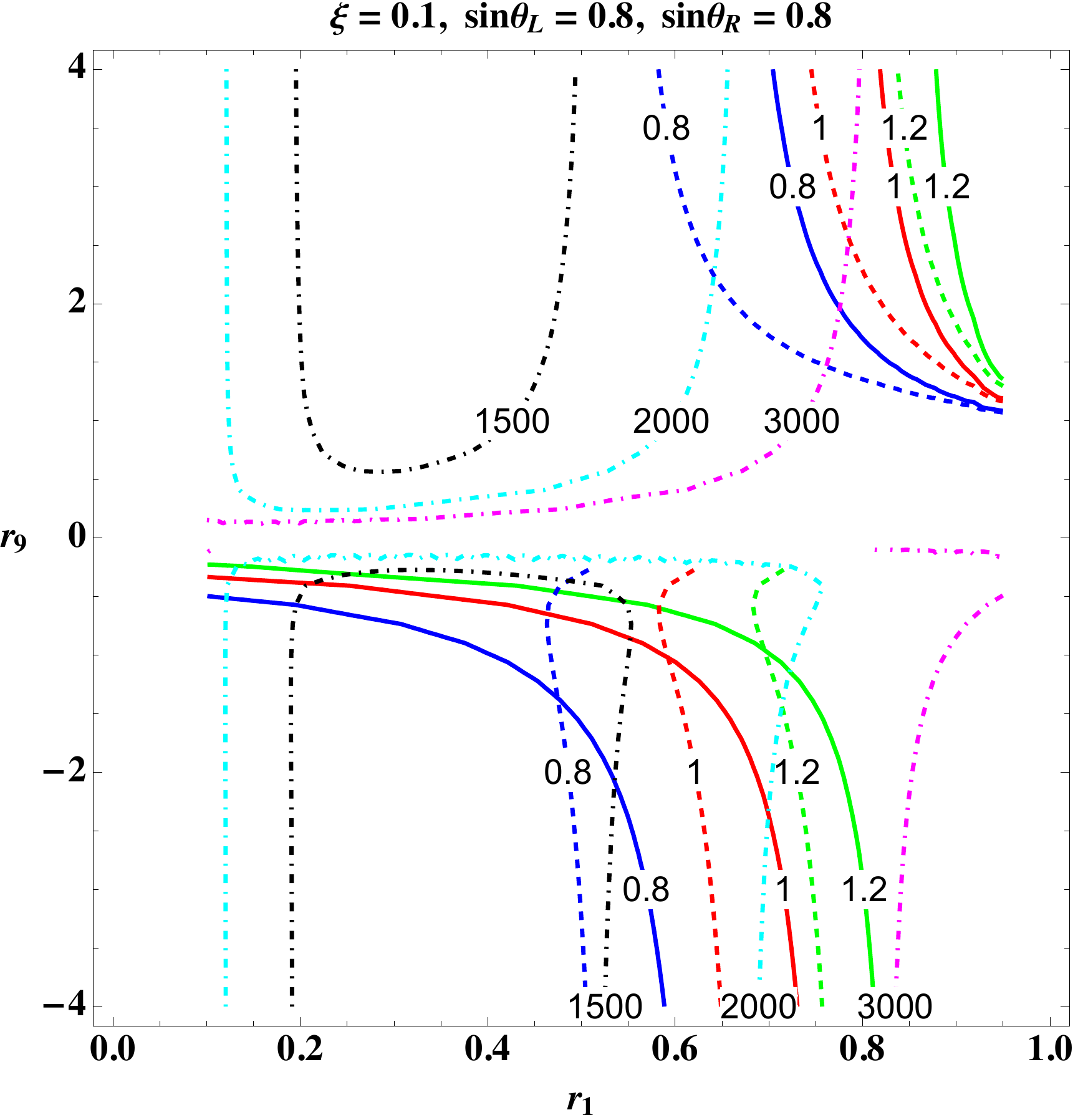}
\includegraphics[width=0.48\textwidth]{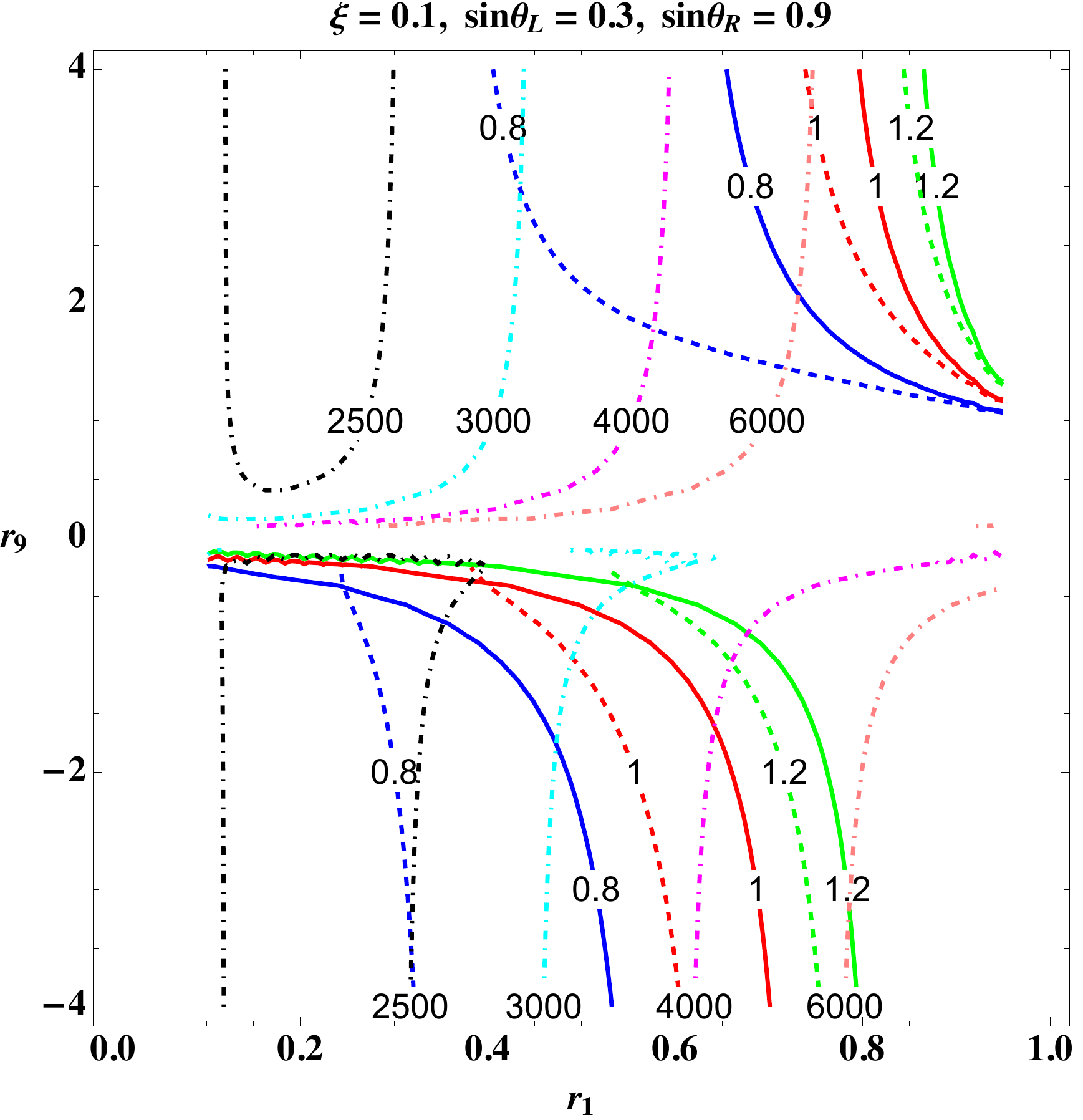}
\end{center}
\caption{The contour plots for $c_g$ (in solid lines) , $c_t$ (in dashed lines)  and $|M_4| [\GeV]$ (in dotted dashed lines). All  plots  use the full formulas in \Eq{eq:cq}, \Eq{eq:cgtform}  and \Eq{eq:cgbform} in the case of $\bold{14}$ and $M_4$ is determined by the top mass $m_t$ = 150 GeV.  }
\label{fig:ctcg14}
\end{figure} 

Next let's turn to  the difference between  the  $t\bar{t}h$ and $ggh$ couplings, which is given by:
\beq
\begin{split} 
\Delta_t - \Delta_g&=\frac14 \sin^2\theta_L  \left(14-\frac{9}{r_9^2} -  \frac{5 }{r_1^2}  \right)  +\frac52\sin^2\theta_R   \left( 1 -  r_1^2 \right)   \\
\end{split}
\eeq
It is evident that:
\beq
\begin{split} 
\Delta_t - \Delta_g& < \frac14 \sin^2\theta_L  \left(14 - \frac{5 }{r_1^2}  \right)  +\frac52\sin^2\theta_R   \left( 1 -  r_1^2 \right)   \\
&<  \frac{7}{2} \sin^2\theta_L  +\frac52\sin^2\theta_R - \frac{5}{\sqrt{2}} |\sin\theta_L \sin\theta_R|
\end{split}
\eeq
from which we can see that:
\beq
\Delta_t - \Delta_g < \frac72
\eeq
Again this quantity enters into the Higgs potential through Eq.~(\ref{eq:gammaff}),
\beq
\label{eq:mfx12}
\mF(x) = \frac{ r_1^2\sec^2\theta_L  \sec^2\theta_R}{(x +\sec^2\theta_L)(x +r_1^2 \sec^2\theta_R)}\left[-\left(\frac1{\xi}(c_t - c_g)+\mO(\xi)\right) \ x +  \mF_0 + \mF_1(x)x\right]
\eeq
where:
\bea
\mF_0 &=& \frac54 \sin^2\theta_L \sin^2 \theta_R (1 - r_1)^2 = \frac{m_t^2}{\xi M_4^2} \left(1+ {\cal O}(\xi) \right) \ ,\\
\mF_1(x) &=&\frac94 \sin^2\theta_L(1 - r_9^2)\left(\frac{\cos^2\theta_R}{r_1^2}-\frac{1}{r_9^2}\right)\left(1 - \frac{r_9^2}{x+ r_9^2} \right)\nonumber \,\\
&&\qquad\qquad+\frac52\sin^2\theta_L \sin^2 \theta_R (1 - r_1^2) \left(1 - \frac{1}{2 \,r_1^2} - \frac12 \frac{1}{x+1}\right) \ .
\label{eq:f1in14}
\eea
Again ${\cal F}(x) >0$ through out the integration region, $0<x<x_\Lambda$, is a sufficient condition to trigger EWSB. Notice that the second contribution to ${\cal F}_1(x)$ in Eq.~(\ref{eq:f1in14}) is identical to ${\cal F}_1(x)$ in the case of $\bold{5}$ in Eq.~(\ref{eq:f1in5})\footnote{Recall that this is the singlet fermion contribution to the $\gamma_f$ in the Higgs potential.}, apart from the numerical coefficient of 5/2. It was shown there that this contribution is quite insignificant when it is positive in the region of $1/2 < r_1^2 < 1$; see Eq.~(\ref{eq:boundf1in5}). The first term in $\mF_1(x)$ is positive in the following region:
\beq
r_9^2 < 1, \qquad r_1^2 < r_9^2 \cos^2\theta_R, \qquad  \text{or} \qquad r_9^2 > 1, \qquad r_1^2 > r_9^2 \cos^2\theta_R
\eeq
and can become sizeable if $|\cos\theta_R/r_1|$ is either much smaller than 1 or much larger than 1. Actually, it is possible to prove:
\bea
\mF_1(x) \leq  \frac94 \sin^2\theta_L \left(1 - \left|\frac{\cos\theta_R}{r_1}\right|\right)^2+\frac52 \left(\frac32 - \sqrt{2}\right)\sin^2\theta_L \sin^2 \theta_R \ .
\eea
In Fig.~\ref{fig:ctcg14}, we perform  numerical scans over $(r_1, r_9,\theta_L,\theta_R)$,  with $M_4$ determined by the 
mass of $m_t$  for $\xi = 0.1, 0.2$. In  the upper plots of $\gamma_f$ versus $c_t - c_g$, we  see that there is a strong preference for $c_t < c_g$ to have a positive $\gamma_f$ to trigger the EWSB. Although there are regions where a significant positive $\gamma_f$ can be obtained for $c_t>c_g$, we show in the lower plot that, once we require that both the value of $\xi$ and the Higgs mass $m_h = 125\GeV$ are reproduced  by the Higgs potential, $c_t$ is always less than $c_g$ in the case of $\xi = 0.1$. Although, we find some points with $c_t > c_g$ in the case of $\xi = 0.2$, both $c_t$ and $c_g$ are very small $< 0.5$ and are not very phenomenologically interesting. Our results in Fig.~\ref{fig:ctcg14} not only confirm the findings in Ref.~\cite{Montull:2013mla}, but also provide an analytic understanding of the strong correlation.


\begin{figure}[!t]
\begin{center}
\includegraphics[width=0.48\textwidth]{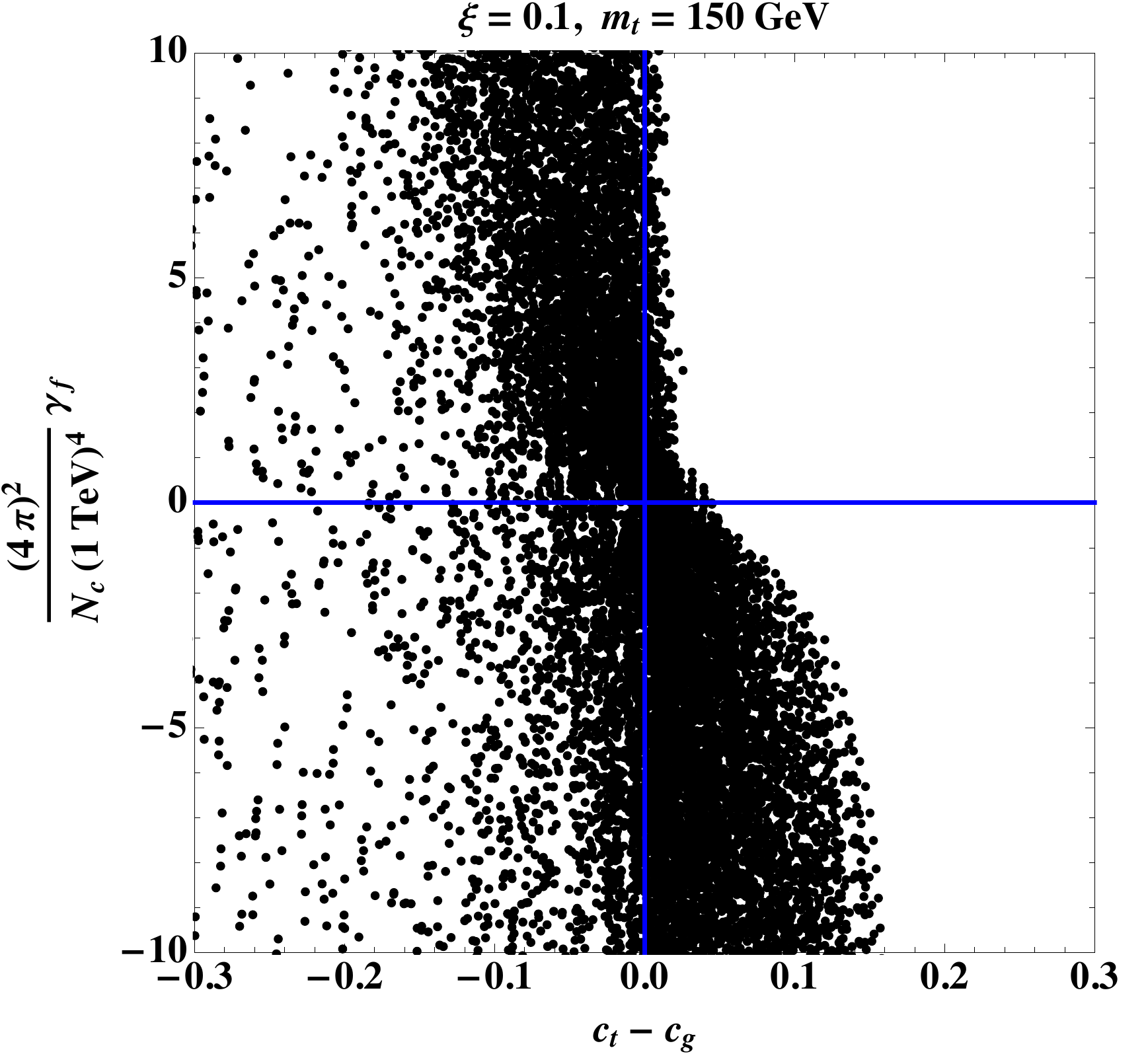}
\includegraphics[width=0.48\textwidth]{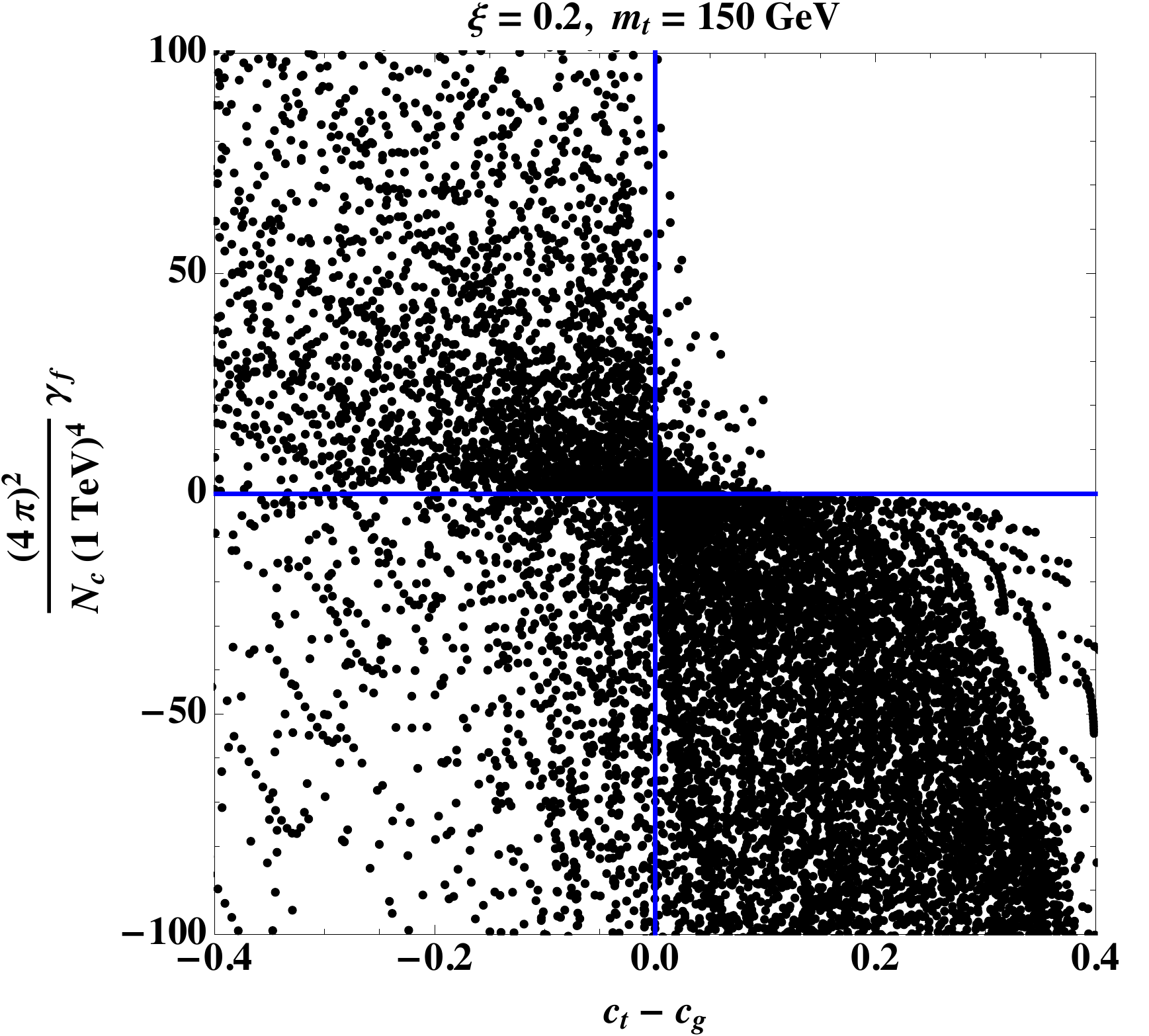}\\
\includegraphics[width=0.48\textwidth]{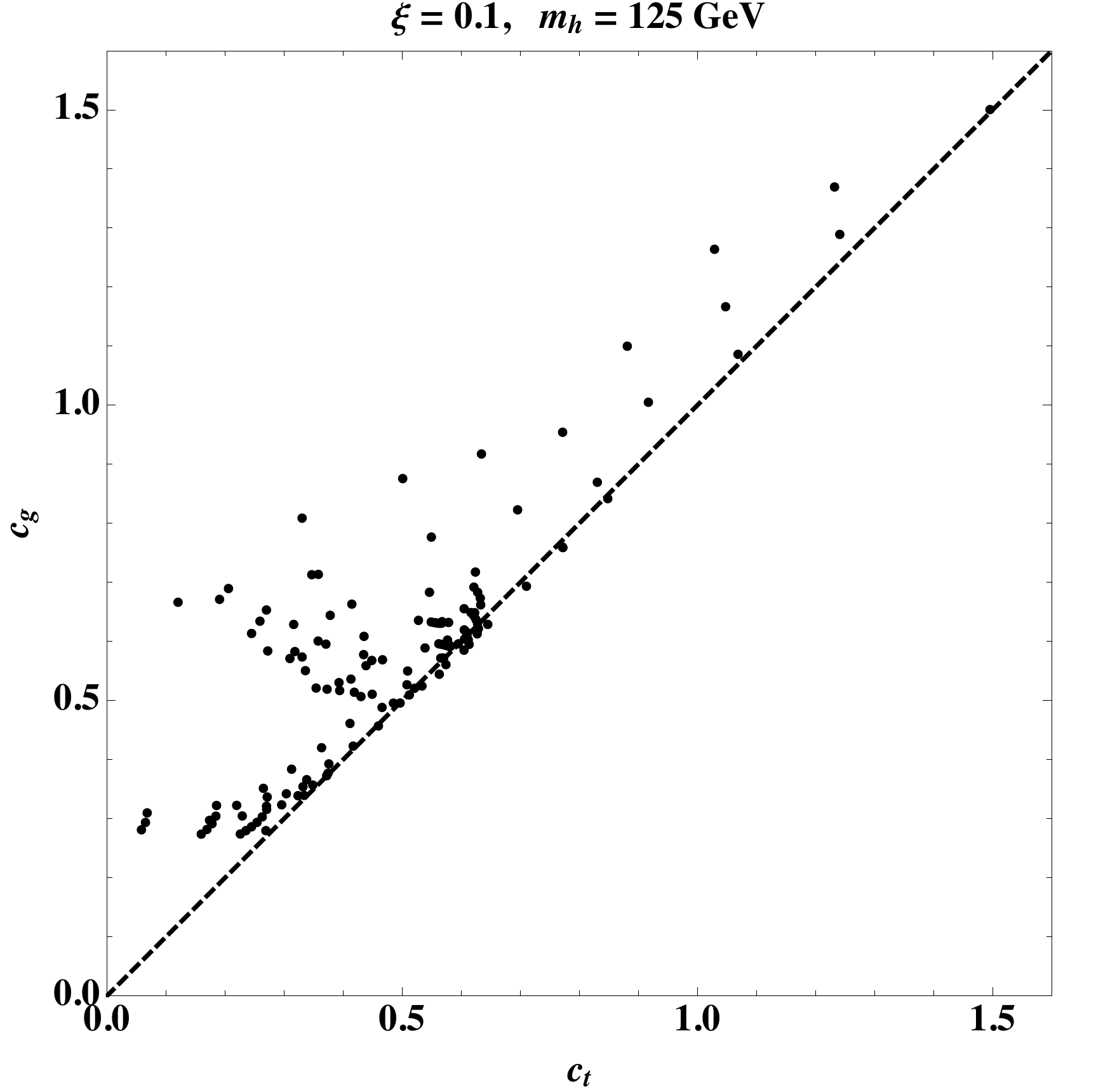}
\includegraphics[width=0.48\textwidth]{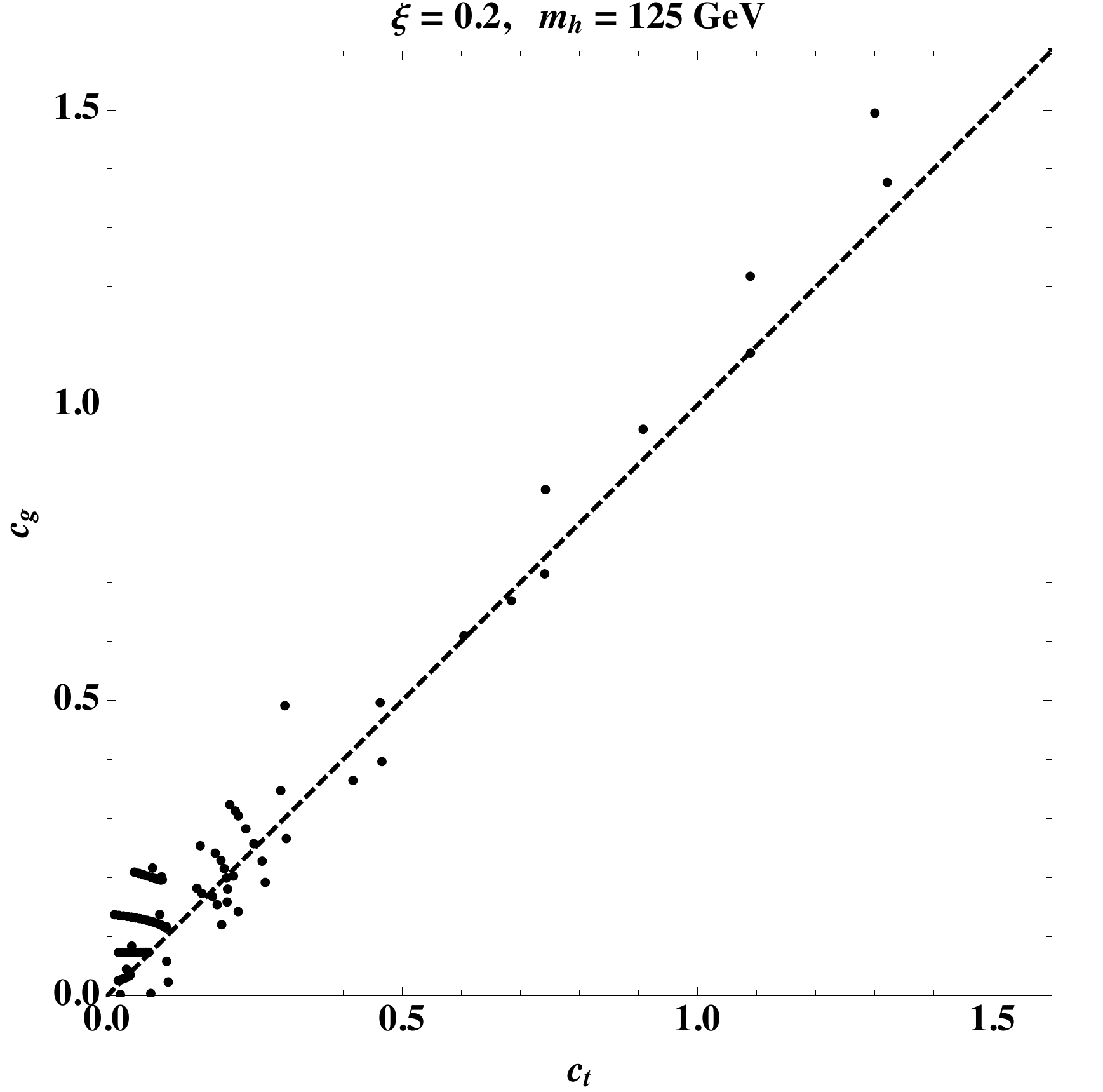}
\end{center}
\caption{ Scattering plots in the case of $\bold{14}$ for $\gamma_f$ versus $c_t - c_g$ (upper panels) without requiring the Higgs potential to reproduce the value of $\xi$ and $m_h$ and $c_t$ versus $c_g$  (lower panels) with $\xi$ and $m_h = 125\GeV$ correctly reproduced for $\xi = 0.1$ (left panels) and $\xi = 0.2$ (right panels). In our the plots, we have required that all the mass scales including the mixing parameters $(c_4 y_L f, a_1 y_R f)$ are smaller than the cutoff $\Lambda = 4 \pi f$ and the lightest top partner is heavier than 500 GeV. }
\label{fig:ctcg14}
\end{figure} 

\subsection{The Bottom Sector}

For the bottom sector, as similar to the case of $\mathbf{5}$ and $\bold{10}$, we introduce composite fermions that mix with $b_R$ but not $t_R$.  The form factors are almost identical to the top except that the mass scales and the mixing parameters are now in the bottom sector. The  elementary doublet $q_L$ are "uplifted" to the $SO(5)$ space via
\beq
q_{L}^{14}  = t_L P_{t_L} + b_L P_{b_L} , \qquad 
b_R^{14} = b_R P_{b_R} \ ,
\eeq
\beq
\begin{split}
(P_{t_L})^{IJ}&= \frac{1}{2} \left(
\begin{array}{cccccc}
&&&&-i\\
&&&&1\\
&&&& 0  \\
&&&&0 \\
-i& 1 &0  & 0&
\end{array} \right), \qquad
(P_{b_L})^{IJ}= \frac{1}{2} \left(
\begin{array}{cccccc}
&&&&0\\
&&&&0\\
&&&& i  \\
&&&& 1\\
0 & 0 & i  & 1&
\end{array} \right), \\
(P_{b_R})^{IJ}&= \frac{1}{2\sqrt{5}} \left(
\begin{array}{cccccc}
-1&&&&\\
&-1&&&\\
&&-1&&   \\
&&&-1&  \\
&&&&4
\end{array} \right) \ .
\end{split}
\eeq
The $SO(5)$ invariants in this case are quite similar to those in the top sector, 
\beq
\begin{split}
 \Sigma^T P_{b_L}^\dagger P_{b_R} \Sigma^* =&  \frac{3}{4\sqrt{5}} s_h c_h\ , \qquad  \Sigma^T P^\dagger_{b_L} \Sigma  \  \Sigma^\dagger P_{b_R} \Sigma^*   =  \left( \frac{2 \sqrt{5}}{5} -   \frac{\sqrt{5}}{2} s_h^2\right) s_h c_h\ , \\
\Sigma^T  P_{b_L}^\dagger P_{b_L} \Sigma^*& = \frac12 - \frac14 s_h^2\ , \qquad \Sigma^T P^\dagger_{b_L} \Sigma \    \Sigma^\dagger P_{b_L} \Sigma^*  = s_h^2 - s_h^4\ ,\\
 \Sigma^T  P_{b_R}^\dagger P_{b_R}  \Sigma^* &  = \frac45 - \frac34 s_h^2\ , \qquad \Sigma^T P^\dagger_{b_R} \Sigma \ \Sigma^\dagger P_{b_R} \Sigma^* = \frac45 - 2 s_h^2 + \frac54 s_h^4\ ,
  \end{split}
\eeq
where the expansions in $s_h^2$ are evident and terminate at the $s_h^4$ order. 
The effective Lagrangian is similar to Eqs.~(\ref{eq:Lagtop14}) to (\ref{eq:Lagtop143}), with $t_R^{14}$ replaced by $b_R^{14}$.

Mixing parameters are defined similar to Eqs.~(\ref{eq:thetaLR14}) and (\ref{eq:r1r914}), with all the parameters now referring to  the bottom partners now. As such we still need $\sin\theta^{(b)}_L\ \sin\theta_R^{(b)} \sim 0.02$ in order to reproduce the small bottom quark mass for  $\xi = 0.1$ and $M_4^{(b)} = 1$ TeV. Therefore their contribution to the Higgs potential can be ignored.

\begin{figure}[tb]
\begin{center}
\includegraphics[width=0.5\textwidth]{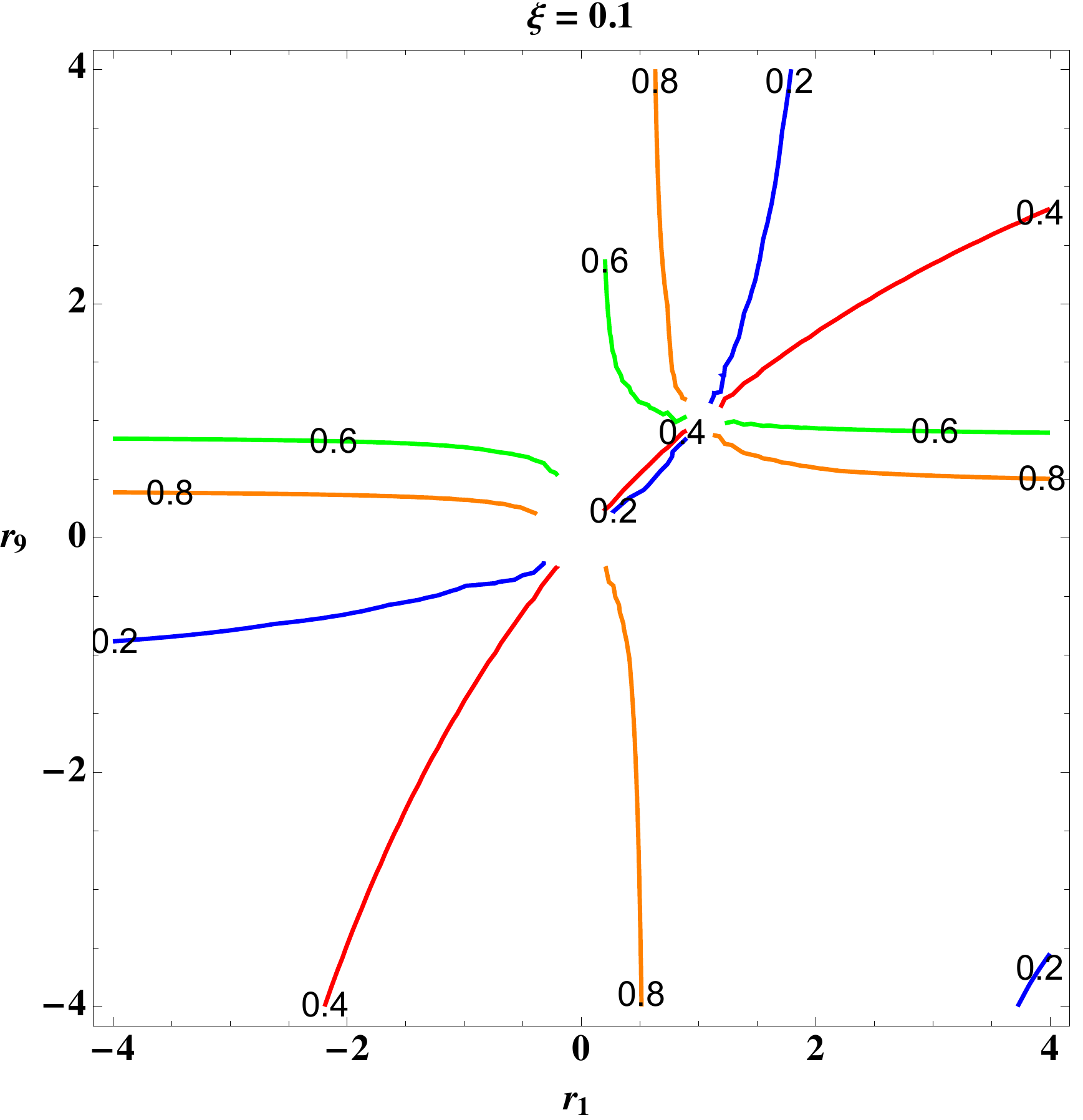}
\end{center}
\caption{Contour plots for the $c_b$ with $\xi = 0.1$, where we have neglected all the mixing parameters in the bottom sector.}
\label{fig:cb14}
\end{figure} 

The modification to the bottom Yukawa coupling is identical to $\Delta_t$ in Eq.~(\ref{eq:Deltat14}), with all the parameters referring to their counterpart in the bottom sector. On the other hand, the $ggh$ coupling from the bottom sector is given by 
\beq
 \Delta_g^{(b)}    =  -\frac52\sin^2\theta^{(b)}_L \left(1 - \frac12\frac{1}{(r_9^{(b)})^2} -  \frac12\frac{1}{(r_1^{(b)})^2}  \right)  -\frac52\sin^2\theta^{(b)}_R   \left( 1 -   (r_1^{(b)})^2 \right) \ ,
\eeq
which is quite different from Eq.~(\ref{eq:Deltag14}) because we have to subtract the SM bottom contribution. In Fig.~\ref{fig:cb14}, we plot the contours of $c_b$ with $\xi = 0.1$ neglecting all the mixing parameters in the bottom sector. Due to the dependence on ratios of  the mass scales, $c_b$ can be suppressed significantly. A phenomenological fit to current Higgs data at the LHC is left for future work.

\section{Conclusions}
\label{conclusions}

In this work we studied patterns of modifications in the Higgs couplings to third generation quarks and to two gluons in the $SO(5)/SO(4)$ minimal composite Higgs model. These three couplings play crucial roles in determining phenomenology of the Higgs boson at the LHC. We first presented a general framework for computing the aforementioned couplings by integrating out, at the leading order, partners of the third generation quarks. Then we applied the computation to three scenarios where the composite fermions are embedded in $\bold{5}$, $\bold{10}$ and $\bold{14}$ of $SO(5)$. We also computed the contribution to the Coleman-Weinberg potential of the Higgs boson from the top sector and demonstrated a strong correlation between the $t\bar{t}h$ and the $ggh$ couplings regions of parameter space where the electroweak symmetry breaking is triggered radiatively by the top sector.

Our findings are summarized in Table \ref{tab:summary}. The interesting patterns are
\begin{itemize}

\item  the $t\bar{t}h$ and $b\bar{b}h$ couplings are always reduced relative to their SM expectations if the composite fermions are embedded in $\bold{5}$ or $\bold{10}$, while these couplings can be either enhanced or suppressed in $\bold{14}$ or $\bold{5}+\bold{10}$.

\item the $ggh$ coupling is always suppressed and independent of the mass scale of the top partner in $\bold{5}$. Such a pattern does not hold for embeddings in other representations.

\item There exists strong correlations between $c_t$ and $c_g$, assuming the top sector gives the dominant contribution to the Coleman-Weinberg potential of the Higgs boson. In regions of parameter space where the electroweak symmetry is triggered, $c_t < c_g$ is strongly preferred. Since the SM gauge boson contributions to the Higgs potential will always tend to preserve the electroweak symmetry, including  them will make the preference even stronger. 

\end{itemize}
These patterns could serve as diagnostic tools should a significant deviation appear in future Higgs measurements. In the absence of deviations, they can be used to potentially constrain the size of $\xi = v^2/f^2$ in the $SO(5)/SO(4)$ composite Higgs models.

It should be emphasized that we have only considered the minimal coset structure of $SO(5)/SO(4)$. Obviously minimality is not always the best guideline when it comes to Nature. In particular, we have assumed through out this work that the dominant contribution to the Coleman-Weinberg potential arises from the top sector. It is conceivable that, if one introduces an additional contribution to the Higgs potential to trigger the EWSB, one could then weakened the strong correlation between $c_t$ and $c_g$. One such possibility is to enlarge the coset structure to include an $U(1)_A$ gauge boson, like in the original Georgi-Kaplan model in Refs.~\cite{Kaplan:1983fs, Kaplan:1983sm}. We hope to return to such a scenario in the future.

\begin{table}[t]
\begin{center}
\begin{tabular}{|c|c|c|c|c|c|c|c|}
\hline
 $\xi=\frac{v^2}{f^2}$  &  $\bold{5}$ & $\bold{10}$ & $\bold{14}$ & $\bold{5} + \bold{10}$  \\
    \hline
$c_g$      &  $1-\frac32\xi$   &   $ 1+\xi\left[- \frac32 + \left(\frac{1}{r_6^2} - 1\right) \sin^2\theta_L\right]$      &  $1+\xi\left[- 4 - \frac32\frac {  1 - 1/r_9 }{1 - 1/r_1 }   +\left(\frac{1}{r_9^2} - 1\right) \sin^2\theta_L\right]$   &   Eq.~(\ref{eq:5and10cg})    \\
\hline
$c_t $ & $< 1- \frac12\xi$ & $ < 1 -\xi $ & no bound& no bound\\
\hline
$c_b $ & $< 1 - \frac12\xi$ & $ <  1 -\xi$ & no bound& no bound\\
\hline
$c_t - c_g$ & $<\xi$ & $ < \frac32 \xi$ & $<\frac72\xi $& $< \xi$\\
\hline
\end{tabular}
\end{center}
\caption{Summary for the leading contribution to $c_g$, $c_t$, $c_b$ and $c_t-c_g$ for the case of $\bold{5},\bold{10},\bold{14},\bold{5} + \bold{10}$, where in $c_g$ we only include the top sector contribution. For the case of $\bold{5} + \bold{10}$, we only consider the case that  only $t_R$ is mixing with both $\bold{5}$ and $\bold{10}$ and $q_L$ is only mixing with $\bold{5}$.}
\label{tab:summary}
\end{table}

\acknowledgements

Work at University of Chicago is supported in part by U.S. Department of Energy grant number DE-FG02-13ER41958. Work at ANL is sup- ported in part by the U.S. Department of Energy under Contract No. DE-AC02-06CH11357. I.L. is supported in part by the U.S. Department of Energy under Contract No. DE-SC0010143.

\appendix

\section{ CCWZ for SO(5)/SO(4)}
\label{app:CCWZ}

We present here our basis for the generators of the $SO(5)$ and $SO(4)$ generators, and review the CCWZ formalism for the $SO(5)/SO(4)$ coset. The $SO(5)$ generators are defined as:
\bea
T^{\hat{a}}_{IJ} &=& - \frac{i}{\sqrt{2}}(\delta^{\hat{a} I} \delta^{5J} - \delta^{\hat{a}J} \delta^{5I})  \,,  \\
T^{a L/R}_{IJ} &=& -\frac{i}{2} \left( \frac12 \epsilon^{abc}(\delta^{bI}\delta^{cJ} -\delta^{bJ}\delta^{cI} )\pm (\delta^{aI}\delta^{4J} - \delta^{aJ} \delta^{4I}  )\right)\nonumber \\
&=& -\frac{i}{2} \left(  \epsilon^{abc} \delta^{bI}\delta^{cJ} \pm (\delta^{aI}\delta^{4J} - \delta^{aJ} \delta^{4I}  )\right)\,.
\label{app:tlr}
\eea
where $\hat{a} = 1,\cdots, 4,$ $a, b,c = 1,2,3 $ and $I,J = 1, \cdots$,5. In Eq.~(\ref{app:tlr}) the $T^L$ ($T^R$) generators take the plus (minus) sign. The generators satisfy $\text{Tr}\ T^AT^B = \delta ^{AB}$ as well as the following commutation relations:
\beq
[T^{aL}, T^{bL}] = i \epsilon^{abc} T^{cL},\quad [T^{aR}, T^{bR}] = i \epsilon^{abc} T^{cR},\quad [T^{aL}, T^{bR}] = 0, \quad
[T^a, T^{\hat{a}}] = t^a_{\hat{b}\hat{a}} T^{\hat{b}} \ .
\eeq
The generators $T^{a\, L}$ and $T^{a\, R}$ correspond to the unbroken $SO(4)\simeq SU(2)_L \times SU(2)_R$ generators.

The Goldstone matrix  is defined as $U = e^{i \Pi}$, where the Goldstone fields $\Pi$ are given by:
 \beq
 \label{eq:Uuni}
\Pi = \frac{\sqrt{2} h^{\hat{a}} }{f}T^{\hat{a}} = \frac{-i}{f} \left(\begin{array}{cc}
0 &   \vec{h} \\
- \vec{h}^T  &0
\end{array}
\right)
\eeq
where the  factor $\sqrt{2}$  is just a convention and can be absorbed by the redefinition of decay constant $f$. The fourplet  $\vec{h}$ can be related with the doublet notation $H$ as follows:
\beq
\begin{split}
\vec{h} &= \frac{1}{\sqrt{2}} \left(\begin{array}{c}
- i (h_u - h^\dagger_u)\\
h_u + h^\dagger_u\\
i (h_d - h^\dagger_d)\\
h_d + h_d^\dagger
\end{array}
\right) , \qquad  H  = \left(
\begin{array}{c}
h_u \\
h_d
\end{array}
\right)
\end{split}
\eeq
from which, we can see that the fourth component of $\vec{h}$ will be our physical Higgs boson in the unitary gauge.
 The matrix $U$  transforms under the non-linearly realized $SO(5)$ as follows:
\be
\label{eq:Utran}
U \to  g \,U \, h^\dagger(x)
\ee
where $g\in SO(5), h(x) \in SO(4)$. The CCWZ covariant objects $d_\mu, E_\mu$ are defined as:
\beq
\begin{split}
-i U^\dagger D_\mu U &=- i U^\dagger \partial_\mu U + U^\dagger A_\mu U =   d^{\hat{a}}_\mu T^{\hat{a}} + E_\mu^a T^a = d_\mu + E_\mu\\
\end{split}
\label{eq:Umatrix}
\eeq
which transform under the non-linearly realized $SO(5)$:
\beq
\label{eq:dandE}
\begin{split}
d_\mu & \rightarrow h(x) \ d_\mu\  h(x)^\dagger \\
E_\mu & \rightarrow h(x)\ E_\mu \ h(x)^\dagger - i h(x)\partial_\mu h(x)^\dagger
\end{split}
\eeq
Note that $E_\mu$  transforms like a gauge field under  $SO(5)$.  The leading two-derivative  effective Lagrangian for the Goldstone bosons is 
\beq
\label{eq:leadingLag}
\frac14 f^2 (d_\mu^{\hat{a}})^2 = \frac12 (\partial_\mu h)^2 + \frac12\frac{f^2\sin^2(\theta + h/f)}{4} ( W_\mu^a -  \delta^{a3}B_\mu)^2
\eeq
where $\theta = \langle h\rangle /f$. By computing the $W$ boson mass we see $v= f\sin\theta = $ 246 GeV. In addition the $hWW$ and $hZZ$ couplings are given by
 \beq
  c_W = g_{hWW}/(g_{hWW})_{SM} = \sqrt{1 -\xi}, \qquad
  c_Z = g_{hZZ}/(g_{hZZ})_{SM}=  \sqrt{1-\xi} \ ,
\eeq
where $\xi = v^2/f^2$. For convenience we  use a non-canonical basis for the gauge bosons, which are not relevant in our analysis.

\section{Partial Compositeness and The $ggh$ coupling}
\label{app:ggh}

In this appendix, we review the general formula for the $ggh$ coupling induced  by the composite resonances under the assumption of {\em partial compositeness}. We are working in the fermion  mass eigenstates, whose interactions with Higgs are parametrized as (see Ref.~\cite{Furlan:2011uq,Azatov:2011qy}):
\beq
-\Delta \mathcal{L} =  \sum_i M_i(\left<h\right>) \bar{\psi}_i \psi_i + \sum_{ij} Y_{ij}\bar{\psi}_i \psi_j h 
\eeq
where $\left<h\right>$ is the vacuum expectation value of the neutral Higgs boson and  $Y$ is a Hermitian matrix assumed to be real and  symmetric.  The partonic cross section for $gg\rightarrow h$ can be obtained by direct calculation \cite{Ellis:1975ap}:
\beq
\hat{\sigma}_{gg\rightarrow h} = \frac{\alpha_s^2 m_h^2}{576\pi}\left|\sum_i \frac{Y_{ii}}{M_i} A_{1/2}(\tau_i)\right|^2 \delta(\hat{s} - m_h^2)
\eeq
where $\tau_i \equiv m_h^2/( 4 M_i^2)$ and fermion loop function $A_{1/2}(\tau)$ is defined as 
\bea
A_{1/2}(\tau) &=& \frac32 \left[\tau + (\tau-1)f(\tau)\right]\tau^{-2}\ , \\
 f(\tau) &=& \left\{\begin{array}{cc}
[\arcsin\sqrt{\tau}]^2, \qquad (\tau \leq 1), \\
-\frac14 \left[\log \frac{1 + \sqrt{1 - \tau^{-1}}}{1-\sqrt{1 - \tau^{-1}}} - i\pi\right]^2, \qquad (\tau > 1)
\end{array}
\right.
\eea
and goes to 1 in the limit of $\tau\rightarrow 0$:
\beq
A_{1/2}(\tau\rightarrow 0) \rightarrow 1.
\eeq
We  define the $ggh$ coupling as:
\beq
g_{ggh} = \sum_i \frac{Y_{ii}}{M_i} A_{1/2}(\tau_i) .
\eeq
It is a very good approximation to include only heavy quarks with masses heavier than the Higgs boson in the loop function. In other words, the $ggh$ effective coupling is  dominated by the sum over $Y_{ii}/M_i$ with $M_i \gg m_h $ \cite{Falkowski:2007hz}:
\bea
\sum_{M_i > m_h} \frac{Y_{ii}}{M_i}& =& \sum_{i} \frac{Y_{ii}}{M_i}  - \sum_{M_i < m_h} \frac{Y_{ii}}{M_i}\nonumber\\
&=& \frac12\frac{\partial}{\partial \left<h\right> } \log \Det (M^\dagger M)- \sum_{M_i < m_h} \frac{Y_{ii}}{M_i}  \ ,
\eea
where we have used $ Y_{ii} = \partial M_i /\partial \left<h\right>$. Note that in our convention, the SM  quark Yukawa coupling satisfies 
$(Y_q/M_q)_{SM} = 1/v$, where $v = 246\GeV$ is the SM Higgs VEV.  Since we work in the large top mass limit and neglect all the lighter quark contributions in the SM, we  have $(g_{ ggh})_{\SM}  = 1/v$. Throughout this work we define the ratio
\beq
\begin{split}
c_g &= \frac{g_{hgg}}{(g_{hgg})_{\SM} } = \frac v2 \frac{\partial}{\partial \left<h\right>} \log  \Det (M^\dagger M) - v \sum_{M_i < m_h} \frac{Y_{ii}}{M_i} \ .
\end{split}
\eeq
 For the charge 2/3 particles including the top and assuming the mass matrix is real the result simply reads
\beq
\label{eq:cgt}
c_g^{(2/3)} = \sin\theta  \frac{\partial}{\partial\theta} \log  \Det\  M_{2/3}\ ,
\eeq
where $v = f \sin\theta=$ and $\theta = \left<h\right>/f$. 
For  the charge -1/3 particles, we need to to extract the contribution from the SM bottom quark, 
\beq
\label{eq:cgb}
\begin{split}
c_g^{(-1/3)}&=  \sin\theta  \frac{\partial}{\partial\theta} (\log\Det\ M_{-1/3}  - \log m_b)\ .
\end{split}
\eeq

\section{Effective Lagrangian After Integrating Out The Top Partners}
\label{app:effLag}

In this appendix we present the formula for the effective Lagrangian obtained from integrating out the composite fields, for the SM quark  ($q_L, t_R, b_R$). The Lagrangian in the momentum space for the composite partners and the mixing terms can be generally written as:
\beq
\mL = \bar{\Psi}_i ( \slashed{p} - M_{\Psi})\Psi^i  + K^\dagger_i \Psi^i + \bar{\Psi}_i K^i
\eeq
where we have neglected all gauge interactions.  Here $i$ generally denotes the indices of the irreducible representations of the unbroken $SO(4)$ and  $K^i$  is constructed with the Goldstone matrix $U$ and the  elementary quark fields ($q_L, t_R, b_R$). The effective Lagrangian after integrating out the top partners by using the equation of motion is simply:
\beq
\label{eq:effgeneral}
\mL_{eff} = - \frac{K^\dagger_i (\slashed{p} + M_{\Psi}) K^i}{p^2 - M_\Psi^2}\ .
\eeq
Note that for two-indices tensor representation of $SO(4)$, the indices should be contracted as $K^\dagger_{ij}\cdots K^{ji}$, which only have effects for anti-symmetric representation.

\section{$\bold{5}+\bold{10}$ }
\label{append:5+10}

In this appendix, we  discuss succinctly  the case where the elementary quark fields $q = (t_L, b_L, t_R, b_R)$  are mixing with both $\bold{5}$ and $\bold{10}$ at the same time.  We find there is a new $SO(5)$ invariant $P^{5\dagger}_{q}  P^{10}_{q} \Sigma^{*}$ and, as a consequence, the $ggh$ and the quark Yukawa couplings are both dependent on the ratios of the composite scales as in the case of $\bold{14}$. 


In the top sector, we will consider the scenario where $t_R$ mixes with composite resonances from both the $\bold{5}$ and the $\bold{10}$, while $q_L = (t_L,b_L)^T$  only mixes with those in the $\bold{5}$. There are other possibilities for the mixing pattern which will not be pursued here. The effective Lagrangian is
\begin{equation}
\label{eq:Lag510t}
\begin{split}
\mathcal{L}^{M1_5} &=   \bar{\Psi}_{\bold{1}} (\slashed{p} - M_1) \Psi_{\bold{1}} +\left[  c_1 y_L f (\bar{q}^5_L)_I U^I_{\ 5} \Psi_{\bold{1}R} +  \, a_1 y_R f (\bar{t}^5_R)_I U^I_{\ 5} \Psi_{\bold{1}L} + h.c.\right]   \\
\mathcal{L}^{M6_{10}} & =  \bar{\Psi}_{\bold{6}} (\slashed{p} - M_6) \Psi_{\bold{6}}+\left[   a_6 y_R  f (\bar{t}^{10}_R)_{IJ} U^I_{\  i} U^J_{\ j} \Psi^{ij}_{\bold{6}L}  + h.c.\right]   \\
\mathcal{L}^{M4_5} &=  \bar{\Psi}_{\bold{4}} (\slashed{p} - M_4) \Psi_{\bold{4}} + \cos\theta_c\left[   c_4 y_L  f (\bar{q}^5_L)_I U^I_{\ i}\Psi^i_{\bold{4}R} +  \, a_4 y_R  f (\bar{t}^5_R)_I U^I_{\  i}\Psi^i_{\bold{4}L}  + h.c.\right]   \\
&- \sin\theta_c\left[   \sqrt{2} \tilde{a}_4 y_R  f (\bar{t}^{10}_R)_{IJ} U^I_{\  i} U^J_{\ 5} \Psi^i_{\bold{4}L}  + h.c.\right] \\
\mathcal{L}^{M4_{10}} & = \bar{\tilde{\Psi}}_{\bold{4}} (\slashed{p}- \tilde{M}_4 ) \tilde{\Psi}_{\bold{4}}  + \sin\theta_c\left[   c_4 y_L  f (\bar{q}^5_L)_I U^I_{\ i}\tilde{\Psi}^i_{\bold{4}R} +  \, a_4 y_R  f (\bar{t}^5_R)_I U^I_{\  i}\tilde{\Psi}^i_{\bold{4}L}  + h.c.\right]   \\
&+ \cos\theta_c\left[  \sqrt{2} \tilde{a}_4 y_R  f (\bar{t}^{10}_R)_{IJ} U^I_{\  i} U^J_{\ 5} \tilde{\Psi}^i_{\bold{4}L}  + h.c.\right] \\
\end{split}
\eeq 
where the relevant projection operators are given in Eqs.~(\ref{eq:projection5}) and (\ref{eq:projection10}) for $\bold{5}$ and $\bold{10}$, respectively. Furthermore, $\theta_c$ is mixing angle between the 4-plet from the $\bold{5}$ and the 4-plet from the $\bold{10}$. In addition to the invariants presented in \Eq{eq:invtop5} and \Eq{eq:invtop10}, we have the following  invariants constructed with embedding vectors $P_q^{5\dagger}, P_q^{10}$:
\beq
\begin{split}
P^{5\dagger}_{t_L}  P^{10}_{t_L} \Sigma^{*} &= P^{5\dagger}_{b_L}  P^{10}_{b_L} \Sigma^{*}  = \frac{c_h}{\sqrt{2}},\quad
P^{5\dagger}_{t_L}  P^{10}_{t_R} \Sigma^{*} = -\frac{s_h}{2\sqrt{2}}, \\
P^{5\dagger}_{t_R}  P^{10}_{t_L} \Sigma^{*} &= \frac{s_h}{2},\qquad \qquad \qquad \quad
P^{5\dagger}_{t_R}  P^{10}_{t_R} \Sigma^{*} = 0 \ .
\end{split}
\eeq
Now it is straightforward to calculate the form factors as before.
After imposing the following condition  for the cancellation of the quadratic divergence,
\beq
\label{eq:c510}
c_4^2 = c_1^2,  \qquad a_4^2 =  a_1^2, \qquad \tilde{a}_4^2 =a_6^2 \ ,
\eeq
the leading modifications to the $ggh$  and $t\bar{t}h$ coupling strengths are given by:
\bea
\label{eq:5and10cg}
\Delta_g^{(t)} &=& - \frac32 +  \frac{\sin\theta_c \cos\theta_c  \left(\frac{M_4}{\tilde{M}_4} -1 \right)}{\sin\theta_c \cos\theta_c \left(\frac{M_4}{\tilde{M}_4} -1 \right)  \pm \sqrt{2}\frac{M_4}{M_1} \pm \sqrt{2}\left(\cos^2\theta_c + \sin^2\theta_c \frac{M_4}{\tilde{M}_4}  \right)  }\ , \\
\label{eq:5and10ct}
\Delta_t &=& \Delta_g^{(t)} +  \frac12 \sin^2\theta_L \left(1 - \frac{1}{r_1^2}\right) + \frac{\tan^2\theta^{(1)}_R}{ 1+ \tan^2\theta^{(1)}_R+\tan^2\theta^{(6)}_R} (1 - r_1^2) \nonumber \\
&&\qquad + \frac12 \frac{\tan^2\theta^{(6)}_R}{ 1+ \tan^2\theta^{(1)}_R+\tan^2\theta^{(6)}_R} \left(1 - \sin^2\theta_c \frac{M_6^2}{M_4^2}-\cos^2\theta_c \frac{M_6^2}{\tilde{M}_4^2} \right) \ ,
\eea
where the two $\pm$ sign in the $\Delta_g^{(t)}$ are not correlated and  we have  also defined:
\beq
\begin{split}
\tan\theta_R^{(1)} &= \frac{a_1 y_R f}{M_1}, \qquad \tan\theta_R^{(6)} = \frac{a_6 y_R f}{M_6}, \qquad r_1^2 = \cos^2\theta_c \frac{M_1^2}{M_4^2} +  \sin^2\theta_c \frac{M_1^2}{\tilde{M}_4^2} \\
\tan^2\theta_L &=  c_4^2 y_L^2 f^2\left(\frac{\cos^2 \theta_c }{M_4^2}+ \frac{\sin^2 \theta_c }{\tilde{M}_4^2}\right)\ .
\end{split}
\eeq

Similar to the case of $\bold{14}$, there is enough degree of freedom such that both $\Delta_g$ and $\Delta_t$ can be positive or negative.
It is not difficult to see that :
\bea
\Delta_t - \Delta_g^{(t)} &<&  \frac12 \sin^2\theta_L \left(1 - \frac{1}{r_1^2}\right) + \frac{\tan^2\theta^{(1)}_R}{ 1+ \tan^2\theta^{(1)}_R+\tan^2\theta^{(6)}_R} (1 - r_1^2) + \frac12 \frac{\tan^2\theta^{(6)}_R}{ 1+ \tan^2\theta^{(1)}_R+\tan^2\theta^{(6)}_R}\nonumber\\
&<& \left(\frac{|\sin\theta_L|}{\sqrt{2}} -  \frac{|\tan\theta^{(1)}_R|}{\sqrt{ 1+ \tan^2\theta^{(1)}_R+\tan^2\theta^{(6)}_R}}\right)^2+ \frac12 \frac{\tan^2\theta^{(6)}_R}{ 1+ \tan^2\theta^{(1)}_R+\tan^2\theta^{(6)}_R}\nonumber\\
&<&\text{Max}\left[  \frac12 \sin^2\theta_L, \frac{\tan^2\theta^{(1)}_R}{ 1+ \tan^2\theta^{(1)}_R+\tan^2\theta^{(6)}_R}\right]+\frac12 \frac{\tan^2\theta^{(6)}_R}{ 1+ \tan^2\theta^{(1)}_R+\tan^2\theta^{(6)}_R} < 1\
\eea


In the bottom sector, we consider the possibilities that  all elementary fields are "uplifted" to  $SO(5)$ multiplets  with $X = 2/3$, which means $b_R$ should be uplifted to an  $SO(5)$ vector with $T^{3R} (b_R) = -1$:
\beq
\begin{split}
(P^{10}_{b_R})^{IJ}&= \frac{1}{2}  \left(
\begin{array}{cccccc}
&&-1&-i& \\
&&i&-1&\\
1&-i&&&  \\
i&1&&& \\
 &  &  &   & 0
\end{array} \right) \ ,
\end{split}
\eeq
and the  embedding vectors for the left-handed fields are the same as the top sector. We consider the case that the left-handed quark doublet $q_L$ is mixing with both $\bold{5}$ and $\bold{10}$. The invariants involving the right-handed bottom quarks are listed as follows:
\beq
\begin{split}
\Sigma^T P_{b_R}^{10\dagger} P^{10}_{b_R} \Sigma^*  = \frac{s_h^2}{2}, \qquad  \Sigma^T P_{b_L}^{10\dagger} P^{10}_{b_R} \Sigma^*  = - \frac{s_h c_h}{2}, \qquad 
P^{5\dagger}_{b_L}  P^{10}_{b_R} \Sigma^{*} = -\frac{s_h}{\sqrt{2}}\ .
\end{split}
\eeq
For simplicity we impose a similar condition to \Eq{eq:c510} to cancel the quadratic divergence in the bottom sector. The modification to the bottom Yukawa coupling is given by
\beq
\Delta_b \sim - \frac32 +  \frac{\sin\theta_c^{(b)} \cos\theta_c^{(b)}  \left(\frac{M_4^{(b)}}{\tilde{M}_4^{(b)}} -1 \right)}{\sin\theta_c^{(b)} \cos\theta_c^{(b)} \left(\frac{M_4^{(b)}}{\tilde{M}_4^{(b)}} -1 \right) \pm \left(\sin^2\theta_c^{(b)} + \cos^2\theta_c^{(b)} \frac{M_4^{(b)}}{\tilde{M}_4^{(b)}}  \right) \pm \frac{M_4^{(b)}}{M_6^{(b)}}}
\eeq
where the two $\pm$ signs are not correlated with each other and we have neglected the mixing angles in the bottom sector. Similar to the case of $\bold{14}$, there is enough degree of freedom to suppress bottom Yukawa coupling as needed.

\bibliographystyle{utphys}
\bibliography{references}

\end{document}